\documentclass[12pt,a4paper,titlepage]{article}
\usepackage{t1enc}
\usepackage{amsmath}
\usepackage{amssymb}
\usepackage{latexsym}
\usepackage{natbib}
\usepackage{epsfig}
\usepackage{textcomp}
\usepackage{alltt}
\usepackage{graphicx}
\usepackage{rotating}
\usepackage{dcolumn}
\usepackage{xypic,t1enc}
\usepackage{float}

\begin{document}
\newcommand{\cip}{\perp\!\!\!\perp}
\newcommand{\nothere}[1]{}
\newcommand{\noi}{\noindent}
\newcommand{\mbf}[1]{\mbox{\boldmath $#1$}}
\newcommand{\cond}{\, |\,}
\newcommand{\hO}[2]{{\cal O}_{#1}^{#2}}
\newcommand{\hF}[2]{{\cal F}_{#1}^{#2}}
\newcommand{\tl}[1]{\tilde{\lambda}_{#1}^T}
\newcommand{\la}[2]{\lambda_{#1}^T(Z^{#2})}
\newcommand{\I}[1]{1_{(#1)}}
\newcommand{\cd}{\mbox{$\stackrel{\mbox{\tiny{\cal D}}}{\rightarrow}$}}
\newcommand{\cp}{\mbox{$\stackrel{\mbox{\tiny{p}}}{\rightarrow}$}}
\newcommand{\cas}{\mbox{$\stackrel{\mbox{\tiny{a.s.}}}{\rightarrow}$}}
\newcommand{\ld}{\mbox{$\; \stackrel{\mbox{\tiny{def}}}{=3D} \; $}}
\newcommand{\nk}{\mbox{$n \rightarrow \infty$}}
\newcommand{\con}{\mbox{$\rightarrow $}}
\newcommand{\dprime}{\mbox{$\prime \vspace{-1 mm} \prime$}}
\newcommand{\Borel}{\mbox{${\cal B}$}}
\newcommand{\bevis}{\mbox{$\underline{\em{Proof}}$}}
\newcommand{\Rd}[1]{\mbox{${\Re^{#1}}$}}
\newcommand{\il}[1]{{\int_{0}^{#1}}}
\newcommand{\pl}[1]{\mbox{\bf {\LARGE #1}}}
\newcommand{\real}{\mathbb{R}}
\newcommand{\bth}{\mbox{\boldmath{$\theta$}}}
\newcommand{\bphi}{\mbox{\boldmath{$\phi$}}}
\newcommand{\bbe}{\mbox{\boldmath{$\beta$}}}
\newcommand{\bbeh}{\mbox{\boldmath{$\hat{\beta}$}}}
\newcommand{\bSig}{\mbox{\boldmath{$\Sigma$}}}
\newcommand{\bLam}{\mbox{\boldmath{$\Lambda$}}}
\newcommand{\bGam}{\mbox{\boldmath{$\Gamma$}}}
\newcommand{\bOm}{\mbox{\boldmath{$\Omega$}}}
\newcommand{\bdel}{\mbox{\boldmath{$\delta$}}}
\newcommand{\cI}{\mbox{\boldmath{$\mathcal{I}$}}}
\newcommand{\cJ}{\mbox{\boldmath{$\mathcal{J}$}}}
\newcommand{\cY}{\mbox{\boldmath{$\mathcal{Y}$}}}
\newcommand{\xtil}{\tilde{\bX}}
\newcommand{\bbes}{\mbox{\boldmath{\scriptsize $\bbe$}}}
\newcommand{\bths}{\mbox{\boldmath{\scriptsize $\bth$}}}
\newcommand{\bX}{{\bf X}}
\newcommand{\bC}{{\bf C}}
\newcommand{\bB}{{\bf B}}
\newcommand{\bD}{{\bf D}}
\newcommand{\bG}{{\bf G}}
\newcommand{\bV}{{\bf V}}
\newcommand{\bH}{{\bf H}}
\newcommand{\bP}{{\bf P}}
\newcommand{\bQ}{{\bf Q}}
\newcommand{\bI}{{\bf I}}
\newcommand{\bS}{{\bf S}}
\newcommand{\bU}{{\bf U}}
\newcommand{\bfb}{{\bf b}}
\newcommand{\ba}{{\bf a}}
\newcommand{\bA}{{\bf A}}
\newcommand{\bZ}{{\bf Z}}
\newcommand{\bx}{{\bf x}}
\newcommand{\bm}{{\bf m}}
\newcommand{\bY}{{\bf Y}}
\newcommand{\bv}{{\bf v}}
\newcommand{\bR}{{\bf R}}
\newcommand{\bW}{{\bf W}}
\newcommand{\bw}{{\bf w}}
\newcommand{\bwun}{{\bf 1}}
\newcommand{\bzro}{{\bf 0}}
\newcommand{\pr}{\prime}
\newcommand{\half}{1/2}
\newcommand{\halffr}{\frac{1}{2}}
\newcommand{\Vh}{\bV^{\half}}
\newcommand{\Vmh}{\bV^{-\half}}
\newcommand{\frsm}[2]{\mbox{\small $\frac{#1}{#2}$}}
\newcommand{\nrm}[1]{\| #1 \|_\infty}
\newcommand{\Var}{\mbox{Var}}
\newcommand{\Cov}{\mbox{Cov}}
\newcommand{\cX}{\mathcal{X}}
\newcommand{\sumi}{\sum_{i=1}^n}
\newcommand{\Zbw}{\bar{Z}_w}
\newcommand{\intrl}{\int_{-\infty}^{\infty}}
\newcommand{\hB}{\hat{B}_n}
\newcommand{\tB}{\tilde{B}_n}
\newcommand{\chB}{\check{B}}
\newcommand{\cV}{\mathcal{V}}
\newcommand{\tUps}{\tilde{\Upsilon}}

\newtheorem{thm}{Theorem}
\newtheorem{lemma}{Lemma}
\newtheorem{prop}{Proposition}

\begin{center}{\Large{\bf Instrumental variables estimation of exposure effects on a time-to-event
 response using structural cumulative survival models}}
 \end{center}
\vspace{0.2cm}

\begin{center}
{ \large Torben Martinussen}\\
Department of Biostatistics\\
University of Copenhagen\\
\O ster Farimagsgade 5B, 1014 Copenhagen K, Denmark\\
email: tma@sund.ku.dk\\ \vspace{4mm}
{ \large Stijn Vansteelandt} \\
Department of Applied Mathematics, Computer Science and Statistics\\
Ghent University  \\
Krijgslaan 281, S9, B-9000 Gent, Belgium \\
email: stijn.vansteelandt@ugent.be\\ 
 \vspace{4mm}
{ \large Eric J. Tchetgen Tchetgen} \\
Department of  Biostatistics and Epidemiology\\
Harvard School of Public Health\\
677 Huntington Avenue
Kresge, Room 822\\
Boston, Massachusetts 02115, US\\
email: etchetge@hsph.harvard.edu\\
\vspace{4mm}
{ \large David M. Zucker}\\
Department of Statistics\\
Hebrew  University\\
Mount Scopus, 91905 Jerusalem, Israel\\
email: david.zucker@mail.huji.ac.il
\end{center}

\vspace{2cm}

\centerline{\sc Summary}
\noindent

The use of instrumental variables for estimating the effect of an exposure on an outcome is popular in econometrics, and increasingly so in epidemiology. This increasing popularity may be attributed to the natural occurrence of instrumental variables in observational studies that incorporate elements of randomization, either by design or by nature (e.g., random inheritance of genes). Instrumental variables estimation of exposure effects is well established for continuous outcomes and to some extent for binary outcomes. It is, however,  largely lacking for time-to-event outcomes because of complications due to censoring and survivorship bias. In this paper, we make a novel proposal under a class of structural cumulative survival models which parameterize time-varying  effects 
of a point exposure directly on the scale of the survival function; these models are essentially equivalent with a semi-parametric variant of the instrumental variables additive hazards model. We propose a class of recursive instrumental variable estimators for these exposure effects, and derive their large sample properties along with inferential tools. We examine the performance of the proposed method in simulation studies and illustrate it in a Mendelian randomization study to evaluate the effect of diabetes on mortality using data from the Health and Retirement Study.
 We further use the proposed method to investigate potential benefit from breast cancer screening on subsequent breast cancer mortality based on the HIP-study.

\noi
 \vspace{3mm}

\noi

\section{Introduction}

A key concern in most analyses of observational studies is whether sufficient and appropriate adjustment was made for confounding of the association between the considered exposure of interest 
	and outcome. This concern can be mitigated to some extent when data are available on an instrumental variable. This is a variable which is (a) associated with the exposure, (b) has no direct effect on the outcome other than through the exposure, and (c) whose association with the outcome is not confounded by unmeasured variables (see e.g. Hern\'an and Robins, 2006). Condition (a) is empirically verifiable, but conditions (b) and (c) are not. However, condition (c) can sometimes be justified in observational studies that incorporate elements of randomization, either by design or by nature
(Didelez and Sheehan, 2007). The plausibility of condition (b) can sometimes be argued on the basis of design elements (e.g. blinding) or a priori contextual knowledge. 

Instrumental variables have a long tradition in econometrics (e.g., Angrist and Krueger, 2001). They have recently become increasingly popular in epidemiology due to a revival of Mendelian randomization studies (Katan, 1986; Davey-Smith and Ebrahim, 2003). Such studies focus on modifiable exposures  known to be affected by certain genetic variants. They then adopt the notion that an association between these genetic variants and the outcome of interest  (e.g., all-cause mortality) can only be explained by an effect of the  exposure on the outcome. This reasoning presupposes that the genetic variants studied satisfy the aforementioned instrumental variable conditions (Didelez and Sheehan, 2007). That is, they should have no effect on the outcome  (e.g., all-cause mortality) other than by modifying the exposure, which can sometimes be justified based on a biological understanding of the functional genetic mechanism. Moreover, their association with the outcome  should be unconfounded, which is sometimes realistic because of Mendelian randomization: the fact that genes are transferred randomly from parents to their offspring. 

Instrumental variables estimation of exposure effects is well established for continuous outcomes that obey linear models. Two-stage least squares (2SLS) estimation proceeds via two ordinary least squares regressions: regressing the exposure variable on the instrument in the first stage, and next regressing the outcome variable on the predicted exposure value in the second stage. 
This approach presumes that the additive exposure effect is the same at all levels of the unmeasured confounders (Hernan and Robins, 2006), which is  rarely plausible in the analysis of 
event times. The IV-analysis of event times is further complicated because of censoring and the fact that the instrumental variables assumptions, even when valid for the initial study population, 
are typically violated within the risk sets composed of subjects who survive up to a given time. Progress is often made via heuristic adaptations of 2SLS estimation, whereby the second stage regression 
is substituted by a Cox regression (see Tchetgen Tchetgen et al. (2015), and Rassen et al. (2008), Cai et al. (2011) for related approaches for dichotomous outcomes), but these have no formal justification 
outside the limited context of rare events (Tchetgen Tchetgen et al., 2015).

To the best of our knowledge, the first formal IV-approach for the analysis of event times was described in Robins and Tsiatis (1991), who parameterised the exposure effect under a structural accelerated failure time model and developed G-estimation methods for it. Their development is very general, and, in particular, can handle continuous exposures.  However, recurring problems in applications have been the difficulty in finding solutions to the estimating equations and obtaining estimators with good precision. 
This is related to the use of an artificial censoring procedure, where some subjects with observed event times are made censored in the analysis in order to maintain unbiased estimating equations. This procedure may lead to an enormous information loss. Moreover, it leads to non-smooth estimating equations (Joffe et al., 2012), so that even simple models are difficult to fit.
Loeys, Goetghebeur and Vandebosch (2005) proposed an alternative approach based on structural proportional hazards models. Their development does not require the use of recensoring, but is more parametric than that of Robins and Tsiatis (1991) as it requires modeling the exposure distribution. It is moreover limited to settings with a binary instrument and constant exposure at one level of the instrument, which is characteristic of placebo-controlled randomized experiments without contamination. Cuzick et al. (2007) relax this limitation by adopting a principal stratification approach but, like other such approaches (see e.g. Abadie, 2003; Nie, Cheng and Small, 2011), restrict their development to binary exposure and instrumental variables. More recently, Tchetgen Tchetgen et al. (2015) independently demonstrated the validity of two-stage estimation approaches in additive hazard models for event times when the exposure obeys a particular location shift model (see Li, Fine and Brookhart (2015) for a related approach under a more restrictive model; other related approaches are discussed in Tchetgen Tchetgen et al., 2015). 
In this article, we avoid restrictions on the exposure distribution and develop IV-estimators under a semiparametric structural cumulative survival model that is closely related to, but less restrictive than the additive hazard model in Tchetgen Tchetgen et al. (2015) and Li, Fine and Brookhart (2015). The proposed approach is general in that it can handle arbitrary exposures and instrumental variables, and can accommodate adjustment for baseline covariates. It neither requires modelling the exposure distribution 
nor the association between covariates and outcome,
and it naturally deals with administrative censoring and certain forms of dependent censoring. 
Picciotto et al. (2012) studied the different problem of adjusting for time-varying confounding
when estimating the effect of a time-varying exposure on a survival outcome. While we also make
use of the structural cumulative failure time model, we do this for handling the different problem
of estimating the effect of an exposure on a survival outcome in the presence of unobserved confounding using an instrumental
variable. Because of this and the fact that we make use of semi-parametric continuous-time
models, in contrast to Picciotto et al. (2012) who focus on parametric discrete-time models, the
recursive estimators that we propose cannot be immediately compared with those in Picciotto
et al. (2012). A further strength of our paper is that it develops an asymptotic inference for
the proposed recursive estimators; such theory is currently lacking for G-estimators in structural
cumulative failure time models.
The semiparametric estimator that we propose requires only a correct model
 for the conditional mean of the instrumental variable, given covariates, for consistency of the estimated causal effect.
 Besides  deriving its large sample properties we also develop  inferential tools allowing us for instance to investigate for time-changing exposure effect. 
We examine the performance of the proposed method in simulation studies and two empirical studies.
  
\section{Model specification and estimation}

\subsection{Basics}

Our goal is to estimate the effect of an arbitrary exposure $X$ on an event time $\tilde{T}$ under the assumption that $G$ is an instrumental variable, conditional on a covariate set $L$. 
A data-generating mechanism that satisfies this assumption is depicted in the causal diagram (Pearl, 2009) of Figure 1. Here, the instrumental variables assumptions are guaranteed by the absence of a direct effect of $G$ on $\tilde{T}$, and by the absence of effects of the unmeasured confounder $U$ on $G$, and of $G$ on $U$. 
 
\begin{figure}[h!]
$$
\xymatrix{& &&   U \ar[dl]\ar[dr] &\\
G\ar[rr]&  &X  \ar[rr] & &\tilde T\\
& & \ar[ull] L \ar[u] \ar[urr] \ar[uur] & &\\
}
 $$
  \caption{Causal Directed Acyclic Graph. $G$ is is the instrument, $X$ the exposure variable and $\tilde T$ the time-to-event outcome. The potential unmeasured confounders are denoted by $U$, and the observed confounders of the $G$-$\tilde{T}$ association by $L$.}
\end{figure}
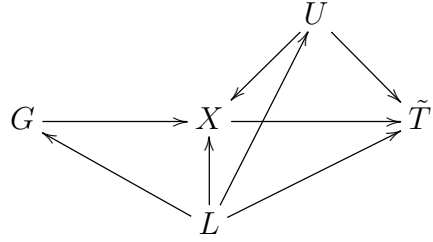


To provide insight, we will start by 
considering uncensored survival data under the following semi-parametric variant of the additive hazards model
(Aalen, 1980):
\begin{equation}
\label{Aalen}
E\left\{d\tilde{N}(t)|\mathcal{F}^{\tilde{N}}_t,G,X,L,U\right\}=\left\{d\Omega(t,L,U)+dB_X(t)X\right\}\tilde{R}(t),
\end{equation}
where $\tilde{N}(t)=I(\tilde{T}\le t)$ denotes the counting process, $\mathcal{F}^{\tilde{N}}_t$ the history spanned by $\tilde{N}(t)$, $\tilde{R}(t)=I(t\le \tilde{T})$ is the at risk indicator, $\Omega(t,L,U)$ is an unknown, non-negative function of time, $L$ and $U$, and 
$B_{X}(t)$ is an unknown scalar at each time $t>0$.
Note that the righthand side of this model does not involve $G$ because of the instrumental variables assumptions, which imply that $\tilde{T}$ and $G$ are conditionally independent, given $X,L$ and $U$. Note furthermore that we explicitly choose to 
leave $\Omega(t,L,U)$ unspecified because $U$ is unmeasured, thus making assumptions about the hazard's dependence on $U$ rather delicate. 

Under model (\ref{Aalen}),
\begin{equation}\label{effect}
\exp\left\{-B_X(t)x\right\}=\frac{P(\tilde{T}>t|X=x,G,L,U)}{P(\tilde{T}>t|X=0,G,L,U)},
\end{equation}
which captures the exposure effect of interest by virtue of conditioning on the unmeasured confounder $U$.
By the collapsibility of the relative risk (or the related collapsibility of the hazard difference (Martinussen and Vansteelandt, 2013)), this is also equal to the directly standardized relative survival risk: 
\[
\exp\left\{-B_X(t)x\right\}=\frac{E\left\{P(\tilde{T}>t|X=x,G,L,U)|G,L\right\}}{E\left\{P(\tilde{T}>t|X=0,G,L,U)|G,L\right\}},
\]
where the averaging is over the conditional distribution of $U$ given $(G,L)$. Letting $\tilde{T}^x$,  for each fixed $x$, denote the potential outcome that would have been observed if the exposure were set to $x$ by some intervention, this can also be written as 
\begin{eqnarray}\label{marginal}
\exp\left\{-B_X(t)x\right\}=\frac{P(\tilde{T}^x>t|G,L)}{P(\tilde{T}^0>t|G,L)}.
\end{eqnarray}
This can be seen because, by definition of $U$ being sufficient to adjust for confounding of the effect of $X$ on $\tilde{T}$, we have that  $\tilde{T}^x$ is conditionally independent of $X$, given $U,G,L$. 

That the effect $\exp\left\{-B_X(t)x\right\}$ can also be defined without making reference to the unmeasured confounder $U$ (that is, without conditioning on $U$) is important. Indeed, the lack of data on $U$ as well as the lack of a precise understanding of the variables contained inside $U$, would otherwise make interpretation difficult (Vansteelandt et al., 2011).

Model (\ref{Aalen}) is closely related to the structural cumulative survival model: 
\begin{equation}
\label{Param}
\exp\left\{-B_X(t)x\right\}=\frac{P(\tilde{T}>t|X=x,G,L)}{P(\tilde{T}^0>t|X=x,G,L)}.
\end{equation}
This model is slightly less restrictive than model (\ref{Aalen}). It makes no assumptions as to how the unmeasured confounders are associated with the event time. It moreover models the effect of setting the exposure to zero, within exposure subgroups rather than the entire population. By evaluating effects within exposure subgroups, the parameter $B_X(t)$ in model  (\ref{Param}) thus encodes a type of treatment effect in the treated. Under the additional assumption that there is no current treatment interaction (Hern\'an and Robins, 2006), a population-averaged interpretation can be made. In particular, suppose that within levels of $G$ and $L$, the effect of exposure level $x$ versus 0 on the survival function is the same for subjects with observed exposure $X=x$ as for subjects with a different exposure level in the following sense 
\begin{equation}
\label{nocurtrtint}
\frac{P(\tilde{T}^x>t|X=x,G,L)}{P(\tilde{T}^0>t|X=x,G,L)}=\frac{P(\tilde{T}^x>t|X\ne x,G,L)}{P(\tilde{T}^0>t|X\ne x,G,L)}.
\end{equation}
Then it is easily verified that, as is the case for model (\ref{Aalen}), model (\ref{Param}) along with the assumption of no current treatment interaction implies (\ref{marginal}), so that $B_X(t)$ captures a population-averaged effect. 

Under the instrumental variables assumptions  that$X$ and $G$ are associated, conditional on $L$, and that $\tilde{T}^0$ is conditionally independent of $G$, given $L$, the estimators of $B_X(t)$ that we will propose in the next section will be consistent estimators of $B_X(t)$ in both models (\ref{Aalen}) and (\ref{Param}).  Condition \eqref{nocurtrtint} is not required for the estimation methods that we develop later on; it is only needed to provide the population-level interpretation given in (\ref{marginal}).

\subsection{Estimation}

We will allow for the event time $\tilde T$ to be subject to right-censoring. In that case, we only observe whether or not $\tilde T$ exceeds a random 
censoring time $C$, i.e. we observe $\delta=I(\tilde T\leq C)$, along with the first time either failure or censoring occurs, i.e. we also observe $T=\min(\tilde T,C)$. Let $(T_i,\delta_i,L_i,G_i,X_i)$, $i=1,\ldots, n,$ denote $n$ independent identically distributed replicates under the structural cumulative failure time model \eqref{Aalen} together with the instrumental variables assumptions. It is assumed that $\tilde T_i$ and $C_i$ are independent given $L_i, G_i, X_i$ and that $P(C_i>t|X_i,G_i,L_i)=P(C_i>t|L_i)$. In fact, the above condition on the censoring distribution can be relaxed to  $P(C>t|X,G,L,U)=P(C>t|L,U)$ for some variable $U\cip G|L$.
The counting processes $N_i(t)=I(T_i\leq t,\delta_i=1)$, $i=1,\ldots, n,$ are observed in the time interval $[0,\tau ]$, where $\tau$ is some finite time point. Further, we define the at risk indicator $R_i(t)=I(t\leq T_i)$, $i=1,\ldots, n$.

The crux of our estimation method for $B_X(t),t>0$ is that once the exposure effect has been eliminated from the event time, it only retains a dependence on $L$ and $U$. It thus becomes conditionally independent of the instrumental variable, given $L$. 
In particular, using arguments similar to those of Martinussen et al. (2011), we eliminate
 the exposure effect from the increments $dN(t)$ by calculating $dN(t)-dB_X(t)X\tilde R(t)$, as suggested by (\ref{Aalen}), and 
we will eliminate the exposure effect from the at risk indicators $R(t)$ by calculating $R(t)\exp\left\{B_X(t)X\right\}$, as suggested by (\ref{effect}). It follows that 
\begin{equation}
\label{EstEq}
E\left[\left\{G-E(G|L)\right\}e^{B_X(t)X}R(t)\left\{dN(t)-dB_X(t)X\right\}\right]=0,
\end{equation}
for each $t>0$, 
which can be seen formally as follows
\begin{eqnarray*}
&&E\left[\left\{G-E(G|L)\right\}e^{B_X(t)X}R(t)\left\{dN(t)-dB_X(t)X\right\}\right]\\
&&=E\left[\left\{G-E(G|L)\right\}e^{B_X(t)X}I(C>t)\tilde{R}(t)d\Omega(t,L,U)\right]\\
&&=E\left[P(C>t|L)\left\{G-E(G|L)\right\}e^{B_X(t)X}\tilde{R}(t)d\Omega(t,L,U)\right]\\
&&=E\left[P(C>t|L)\left\{G-E(G|L)\right\}e^{-\Omega(t,L,U)}d\Omega(t,L,U)\right]\\
&&=0;
\end{eqnarray*}
here, the last equality follows because $G\cip U|L$.

The unbiasedness of equation \eqref{EstEq} suggests a way of estimating the increments $dB_X(t)$ by solving equation \eqref{EstEq} for each $t$ with population expectations replaced by sample analogs. 
%
This delivers  
 the recursive estimator $\hat B_X(t)$ defined by
\begin{equation}
\label{Est}
\hat B_X(t,\hat \theta)=\int_0^t \frac{\sum_iG^c_i(\hat\theta)e^{\hat B_X(s-)X_i}dN_i(s)}{\sum_iG^c_i(\hat\theta)R_i(s)e^{\hat B_X(s-)X_i}X_i},
\end{equation}
where $G^c_i(\theta)=G_i-E(G_i|L_i;\theta)$, with $E(G_i|L_i;\theta)$ a parametric model for $E(G_i|L_i)$ and $\hat\theta$ a consistent estimator of $\theta$ (e.g., a maximum likelihood estimator). 

The estimator (\ref{Est}) is given by a counting process integral, thus only changing values at observed death times. Because of its recursive structure, we calculate it forward in time, starting from $\hat B_X(0)=0$. 
In the special case where the exposure $X$ is binary, it can be 
calculated analytically as
 shown below. With $A(t)=e^{B_X(t)}$, the equation (\ref{EstEq}) (with population expectations substituted by sample averages) leads to
$$
dA(t)=
\frac{\sum_iG^c_i(\theta)(1-X_i)dN_i(t)}{\sum_iR_i(t)G^c_i(\theta)X_i}
+A(t)\frac{\sum_iG^c_i(\theta)X_idN_i(t)}{\sum_iR_i(t)G^c_i(\theta)X_i}
$$
When replacing  $A(t)$ with $A(t-)$ on the right side of this expression and integrating,
we get the Volterra equation (see Andersen et al. (1993), p. 91)
$$
A(t)=W(t)+\int_0^tA(s-)dU(s),
$$
where 
$$W(t)=\int_0^t\frac{\sum_iG^c_i(\theta)(1-X_i)dN_i(s)}{\sum_iR_i(s)G^c_i(\theta)X_i},\quad
U(t)=\int_0^t\frac{\sum_iG^c_i(\theta)X_idN_i(s)}{\sum_iR_i(s)G^c_i(\theta)X_i},$$
 and the solution is given by
$$
A(t)=W(t)+\int_0^t[\prod_{(s,t]}\{1+dU(v)\}]dW(s)
$$

With the additional assumption that $dB_X(t)=\beta_Xdt$, that is assuming a time-constant effect,
an estimator of $\beta_X$ may be obtained as 
\begin{equation}
\label{semi.est}
\hat \beta_X=\int_0^{\tau} w(t)d\hat B_X(t)
\end{equation}
with $w(t)=\tilde w(t)/\int_0^{\tau}\tilde w(s)\, ds$, $\tilde w(t)=R_{\mbf{\cdot}}(t)=\sum_iR_i(t)$.

\section{Large sample results}


The following proposition, whose proof is given in the Appendix, shows that $\hat B_X(t)$  is a uniformly consistent estimator of  $B_{X}(t)$. It moreover gives the asymptotic distribution of $\hat B_X(t)$. 

\begin{prop}
Under model \eqref{Param} with the assumption that $G$ is an instrumental variable, conditional on $L$, and given the technical conditions listed  in the Appendix, the  IV estimator  $\hat B_{X}(t)$ is a uniformly consistent estimator of $B_{X}(t)$. Furthermore,
 $
 W_n(t)=n^{1/2}\{\hat B_{X}(t,\hat \theta)-B_X(t)\}
 $
 converges in distribution to a zero-mean Gaussian process with variance $\Sigma(t)$. A uniformly consistent estimator $\hat\Sigma(t)$ of $\Sigma(t)$ is given below.
\end{prop}
\medskip

Let $\epsilon_i^B(t,\theta),i=1,...,n$ be the iid zero-mean processes given by expression \eqref{Eps} in the Appendix.
From the proof in the Appendix, it then follows that
$W_n(t)$ is asymptotically equivalent to
$n^{-1/2}\sum_{i=1}^n\epsilon_i^B(t,\theta)$.
 The variance $\Sigma(t)$ of the limit distribution can thus be consistently estimated by
$$
\hat \Sigma(t)=n^{-1}\sum_{i=1}^n\{\hat \epsilon_i^B(t,\hat\theta)\}^2,
$$
where $\hat \epsilon_i^B(t,\hat \theta)$ is obtained from $\epsilon_i^B(t,\theta) $ by replacing unknown quantities with their empirical
counterparts.
These results
can be used to construct a pointwise confidence band.
The asymptotic behavior of the estimator \eqref{semi.est} is easily obtained since:
$$
n^{1/2}(\hat \beta_X-\beta_X)=n^{-1/2}\sum_{i=1}^n\int_0^{\tau}w(t)d\epsilon_i^B(t,\theta).
$$

To study temporal changes, it is more useful to consider a uniform confidence band.   This and
tests of the hypothesis of a linear (cumulative) causal effect
$$
\mbox{H}_0:\; \beta_{X}(t)=\beta_{X} \ \ \Leftrightarrow \ \  \mbox{H}_0:\; B_{X}(t)=\beta_{X} t
$$
can easily be derived based on the above iid representation as also outlined in Martinussen (2010).
The above hypothesis can be tested using the following test statistic
\begin{equation}
\label{sup-test}
\sup_{t\leq \tau}|n^{1/2}(\hat B_X(t)-\hat \beta_X t)|,
\end{equation}
and since, under the null, $n^{1/2}(\hat B_X(t)-\hat \beta_X t)=n^{1/2}\{\hat B_X(t)-B_X(t)-(\hat \beta_X - \beta_X)t\}$ it is easy to get the iid-representation of the test process.
This development is based on the following 
approach of Lin et al. (1993).
Let $Q_1^m,\ldots ,Q_n^m$ be independent standard normal variates. Then, given the data,
$$
\hat W_m(t)=n^{-1/2}\sum_{i=1}^n\hat \epsilon_i(t,\hat \theta)Q_i^m
$$
also converges in distribution to a zero-mean Gaussian process with variance $\Sigma(t)$.
 The limit distribution can thus be evaluated by generating a large number, $M$, of replicates $\hat W_m(t)$,
$m=1, \ldots ,M$. The causal null hypothesis that $B_X(t)=0$ for all $t$ can thus for example be tested using the test statistic
$$
\sup_{t\leq \tau}|n^{1/2}\hat B_X(t)|,$$
by investigating how extreme this statistic is in the distribution of $\sup_{t\leq \tau}|\hat W_m(t)|$, $m=1, \ldots ,M$. 

\nothere{
The causal null hypothesis that $B_X(t)=0$ for all $t$ can be tested along the same lines, or can be more easily tested 
by using the above resampling approach with 
\[\hat{\epsilon}_i(t,\hat \theta)=\int_0^t \left[G_i-\mu(L_i;\hat{\theta})-\left\{\frac{1}{n}\sum_{i=1}^n \frac{\partial \mu(L_i;\hat{\theta})}{\partial\theta} R_i(s)dN_i(s)\right\} \epsilon_i^{\theta}\right]\]
where $\mu(L_i;\theta)=E(G_i|L_i;\theta)$, and $\epsilon_i^{\theta}$ is defined in the Appendix. 
}
 

\nothere{
\textit{Stijn: Should we add something about that the additive hazards model will often give a good approx. to any hazard function as it is a first order Taylor expansion? TORBEN, I THINK THAT WOULD BE USEFUL. SHOULD WE DO THAT ALREADY IN THE INTRODUCTION?}

}

\nothere{
\subsection{Efficient estimation}

In the Appendix, we show that, up to asymptotic equivalence, all regular and asymptotically linear estimators of $B_X(t)$ can be obtained by solving an estimating equation with estimating function:
\begin{eqnarray*}
&&\int \left[x_1(s,G,L)-E\left\{x_1(s,G,L)|L\right\}\right]\left[R_0(s)dN_0(s)-E\left\{R_0(s)dN_0(s)|L\right\}\right]\\
&&+\left[x_2(s,G,L)-E\left\{x_2(s,G,L)|L\right\}\right]\left[R_0(s)-E\left\{R_0(s)|L\right\}\right]ds\\
&& + x_3(G,L)-E\left\{x_3(G,L)|L\right\},
\end{eqnarray*}
for specific functions $x_1(s,G,L),x_2(s,G,L)$ and $x_3(G,L)$, where 
\begin{eqnarray*}
R_0(s)&\equiv & R(s)\exp\left\{B_X(s)X\right\}\\
dN_0(s)&\equiv & dN(s)-dB_X(s)X.
\end{eqnarray*}
For given, $x_1(s,G,L)$ and $x_2(s,G,L)$, the most efficient estimator of $B_X(t)$ corresponds with the choice 
$x_3(G,L)=0$. Finding the optimal functions $x_1(s,G,L)$ and $x_2(s,G,L)$ that deliver a semi-parametric efficient estimator, requires solving complex integral equations (see the Appendix). 
We therefore recommend using
\begin{eqnarray*}
&&\int \left[E\left\{X\exp\left(B_X(s)X\right)|G,L\right\}-E\left\{X\exp\left(B_X(s)X\right)|L\right\}\right]\left(R_0(s)dN_0(s)-E\left\{R_0(s)dN_0(s)|L\right\}\right.\\
&&\left.- d\times E\left\{dN_0(s)|R(t)=1,L\right\}\left[R_0(s)-E\left\{R_0(s)|L\right\}\right]\right)ds,
\end{eqnarray*}
with $d=0,1$, which we expect will deliver a reasonably efficient estimator.
}

\section{Numerical results}

\subsection{Simulation study}
To investigate the properties of our proposed methods with practical
sample sizes, we conducted a simulation study.  We generated  data  according to the data-generating mechanism of  Figure 1 with the following specific models where we leave out the covariate $L$ for simplicity. 
We considered two different settings where the exposure variable was  continuous and binary, respectively.
In the first setting
the exposure  variable $X$ was  continuous. We took $G$ to be binary with $P(G=1)=0.5$, and generated $X$ and $U$, given $G$, from a normal distribution with 
$E(X|G=g)=0.5+\gamma_Gg$, $E(U|G=g)=1.5$ and with variance-covariance matrix so that $Var(X|G)=Var(U|G)=0.25$, and $Cov(X,U|G)=-1/6$.
The parameter  $\gamma_G$  determines the size of the correlation between exposure and the instrumental variable. Specifically we looked at correlation $\rho$ equal to 0.3 and 0.5.
We generated $\tilde T$ according to the  hazard model
$$
E\left\{d\tilde N(t)|T\geq t, X,G,U\right\}=\beta_0(t)+\beta_X(t)X+\beta_U(t)U,
$$
with $\beta_0(t)=0.25$, $\beta_X(t)=0.1$ and $\beta_U(t)=0.15$. Twenty percent were potentially censored according to a uniform distribution on (0,3.5), and the rest were censored at $t=3.5$, corresponding to the study being closed at this time point, leading to an cumulative censoring rate of around 20.
Under this model, as seen in the Section 2.2,  \eqref{Param} 
holds with $B_X(t)=\int_0^t\beta_X(s)\; ds=0.1t$. 
Under this model it further holds that 
$$
E\left\{d\tilde N(t)|T\geq t, X,G\right\}=\tilde\beta_0(t)+\tilde\beta_G(t)G+\tilde\beta_X(t)X,
$$
with $\tilde\beta_X(t)=0$ so the naive Aalen estimator (using $X$ and $G$ as covariates) is biased.
We calculated the estimator given in (\ref{Est}) with $\hat \theta=\overline{G}$, along with the estimator $\hat \beta_X$ given in (9) where we took $\tau=3$.
For this scenario, we considered sample sizes 1600 and 3200 when $\rho=0.3$, and sample sizes 800 and 1600 when $\rho=0.5$.
 Simulation results concerning $\hat B_X(t)$, based on 2000 runs for each configuration,
  are given in Table 1, where (average) biases are reported at time points $t=1, 2, 3$ for $\hat B_{X}(t)$ along with coverage probability of 95\% pointwise confidence intervals CP($\hat B_{X}(t)$). Results concerning $\hat \beta_X$ are given in Table 3, first half.

\vspace{0.5cm}
\centerline{Table 1 about here}
\vspace{0.5cm}

\begin{table}
\caption{
Continuous exposure case. Time-constant exposure effect.
Bias of $\hat B_{X}(t)$, average estimated standard error, sd($\hat B_{X}(t)$), empirical standard error, see($\hat B_{X}(t)$)), and coverage probability of 95\% pointwise confidence intervals CP($\hat B_{X}(t)$)) based on the instrumental variables estimator, in function of sample size $n$ and at different strengths $\rho$ (correlation) of the instrumental variable. Bias of $\tilde B_{X}(t)$ is the bias of the naive Aalen estimator.
 \bigskip}

{\small
\begin{center}
\begin{tabular}{|l|cccc|cccc|}
\hline
 &&\multicolumn{3}{c|}{ $\rho=0.3$}&&\multicolumn{3}{c|}{ $\rho=0.5$} \\
&n & $t=1$ & $t=2$ & $t=3$ & n& $t=1$ & $t=2$ & $t=3$\\

\hline
Bias $\hat B_{X}(t)$ &1600&-0.003   &   -0.001     &  -0.007&800 &       -0.002&-0.004   &  - 0.015      \\
sd ($\hat B_{X}(t)$)&&0.139 &0.242& 0.404& &0.109&0.187 &0.303\\
see ($\hat B_{X}(t)$)&&0.139 &0.245& 0.439& &0.107&0.187 &0.314\\
95\% CP($\hat B_{X}(t)$)&&95.4&     96.5&      98.1   &&   95.2 &96.1&     97.5\\
Bias $\tilde B_{X}(t)$ &&-0.101   &   -0.201     &  -0.300& &      -0.099&-0.197   &  - 0.297      \\
Bias $\hat B_{X}(t)$& 3200&-0.003   &   -0.005    &  -0.014&  1600 &    0.004&0.004   &   -0.002      \\
sd ($\hat B_{X}(t)$)&&0.094 &0.170& 0.267& &0.075&0.131 &0.209\\
see ($\hat B_{X}(t)$)&&0.096 &0.166& 0.262&& 0.075&0.130 &0.206\\
95\% CP($\hat B_{X}(t)$)&&95.6&     95.1&      96.2   &&   95.0 &95.5&     95.7 \\
Bias $\tilde B_{X}(t)$ &&-0.099  &   -0.200     &  -0.296&&       -0.099&-0.200   &   -0.301       \\
\hline
\end{tabular}
\end{center}
}
\end{table}
\bigskip
\bigskip

\begin{table}
\caption{
Continuous exposure case. Time-dependent exposure effect.
Bias of $\hat B_{X}(t)$, average estimated standard error, sd($\hat B_{X}(t)$), empirical standard error, see($\hat B_{X}(t)$)), and coverage probability of 95\% pointwise confidence intervals CP($\hat B_{X}(t)$)) based on the instrumental variables estimator, in function of sample size $n$ and at different strengths $\rho$ (correlation) of the instrumental variable. Size of sup-test is the size of the test based on the statistic \eqref{sup-test} using 2000 re-samplings, and taking $\tau=3$.
 \bigskip}

{\small
\begin{center}
\begin{tabular}{|l|cccc|cccc|}
\hline
 &&\multicolumn{3}{c|}{ $\rho=0.3$}&&\multicolumn{3}{c|}{ $\rho=0.5$} \\
&n & $t=1$ & $t=2$ & $t=3$ & n& $t=1$ & $t=2$ & $t=3$\\

\hline
Bias $\hat B_{X}(t)$ &1600&0.005   &   0.008     &  0.001&800 &       -0.001&0.001   &  - 0.006      \\
sd ($\hat B_{X}(t)$)&&0.136 &0.224& 0.336& &0.108&0.176 &0.249\\
see ($\hat B_{X}(t)$)&&0.138 &0.228& 0.363& &0.107&0.176 &0.264\\
95\% CP($\hat B_{X}(t)$)&&96.2&     96.2&      96.5   &&   95.2 &96.0&     97.1\\
\hline
Bias $\hat B_{X}(t)$& 3200&0.003   &   -0.001    &  -0.004&  1600 &    0.001&0.005   &   0.003      \\
sd ($\hat B_{X}(t)$)&&0.097 &0.156& 0.224& &0.076&0.122 &0.175\\
see ($\hat B_{X}(t)$)&&0.096 &0.157& 0.230&& 0.075&0.121 &0.173\\
95\% CP($\hat B_{X}(t)$)&&95.1&     95.4&      96.6   &&   94.8 &95.0&     95.5 \\
\hline
\end{tabular}
\end{center}
}
\end{table}
\bigskip
\bigskip

\nothere{
\begin{table}
\caption{
Continuous exposure case. Time-dependent exposure effect.
Bias of $\hat B_{X}(t)$, average estimated standard error, sd($\hat B_{X}(t)$), empirical standard error, see($\hat B_{X}(t)$)), and coverage probability of 95\% pointwise confidence intervals CP($\hat B_{X}(t)$)) based on the instrumental variables estimator, in function of sample size $n$ and at different strengths $\rho$ (correlation) of the instrumental variable. Bias of $\tilde B_{X}(t)$ is the bias of the naive Aalen estimator.
 \bigskip}

{\small
\begin{center}
\begin{tabular}{|ll|ccc|ccc|}
\hline
$n$& &\multicolumn{3}{c|}{ $\rho=0.3$}&\multicolumn{3}{c|}{ $\rho=0.5$} \\
& & $t=1$ & $t=2$ & $t=3$ & $t=1$ & $t=2$ & $t=3$\\

\hline
800&Bias $\hat B_{X}(t)$ &-0.009   &   -0.011     &  -0.033&       0.002&0.004   &   -0.009      \\
&sd ($\hat B_{X}(t)$)&0.196 &0.353& 0.629& 0.107&0.193 &0.314\\
&see ($\hat B_{X}(t)$)&0.196 &0.350& 0.669& 0.106&0.186 &0.317\\
&95\% CP($\hat B_{X}(t)$)&95.9&     97.3&      98.5   &   95.4 &95.2&     96.5\\
&Bias $\tilde B_{X}(t)$ &-0.101   &   -0.202     &  -0.306&       -0.098&-0.200   &   -0.299      \\
1600&Bias $\hat B_{X}(t)$ &0.005   &   0.000     &  0.017&       -0.001&-0.003   &   0.013      \\
&sd ($\hat B_{X}(t)$)&0.138 &0.242& 0.393& 0.075&0.128 &0.201\\
&see ($\hat B_{X}(t)$)&0.138 &0.240& 0.405& 0.075&0.130 &0.205\\
&95\% CP($\hat B_{X}(t)$)&95.6&     95.9&      97.8   &   95.6 &95.6&     97.0 \\
&Bias $\tilde B_{X}(t)$ &-0.099   &   -0.196     &  -0.296&       -0.100&0.197   &   0.294       \\
\hline
\end{tabular}
\end{center}
}
\end{table}
\bigskip
\bigskip
}

\noindent
In all scenarios considered the naive Aalen estimator is, as expected, biased; see Table 1.
From Table 1 it is  also seen that  the proposed estimator  $\hat B_X(t)$ is unbiased. 
In the case with sample size 800 and correlation equal to 0.3  the estimated standard error at
time point $t=3$ is a bit too large resulting in a too high coverage probability. However, it is also seen that
the estimated standard error approaches the empirical standard deviation
 as sample size goes up, and overall the 95\%-coverage probabilities have the correct size. We also calculated the size of the sup-test \eqref{sup-test} that investigates whether the constant exposure effects model is acceptable. For the four considered scenarios of $(n,\rho)$: (1600,0.3), (3200,0.3), (800,0.5), (1600,0.5), it was 0.03, 0.04, 0.03 and 0.05, respectively. Hence, when sample size and correlation goes up, the test has the correct size. 
The results concerning the constant effect estimator, $\hat\beta_X$, are reported in the first half of Table 4, and from there it is seen that the estimator is unbiased and that the variability is well estimated leading to satisfactory coverage probabilities at least when sample size goes up. When the exposure is continuous one may also calculate the 2SLS estimator of Tchetgen et al. (2015), we denote it $\check\beta_X$. Results for this estimator are also given in Table 4. From there it is seen that this estimator is also unbiased, and that it is sligtly more efficient than the constant effects estimator given in this paper. This is not surprising as the 2SLS estimator is targeted at this specific situation while the estimator $\hat\beta_X$ is derived from an estimator that can handle much more general situations.
We also considered a setup where there was a time-varying exposure effect. Data was generated as described above except that $\beta_X(t)$ was now taken as 
$\beta_X(t)=0.1I(t<1.5)-0.1I(1.5\le t<3)$. Inducing censoring as above resulted in a cumulative censoring rate of around 25. Results from this study are given in Table 2, where we have dropped results for the naive Aalen estimator. From Table 2 we see again that the proposed estimator is unbiased and that the variability is well estimated resulting in appropriate coverage.
We also calculated the size of the sup-test. For the four considered scenarios of $(n,\rho)$: (1600,0.3), (3200,0.3), (800,0.5), (1600,0.5), it was 0.07, 0.18, 0.13 and 0.31, respectively.  We also ran the situation where $(n=3200,\rho=0.5)$ and obtained  the size of the test to be 0.61.
 Whe thus see, as expected, that when correlation and sample size goes up the power of the test increases. We also calculated the constant effects estimators $\hat\beta_X$ and $\check\beta_X$, and the mean of them in all four combinations of $(n,\rho)$ was 0.04 thus showing that
 the constant effects estimators are not appropriate under this scenario with time-changing exposure effect.

We also considered  settings where the exposure variable $X$ was binary. In the first such setting we generated data as under the first scenario with $\beta_X(t)=0.1$, but instead of using the continuous version of $X$, call it now $\tilde X$, we used $X=I(\tilde X>0.5)$. 

\vspace{0.5cm}
\centerline{Table 2 about here}
\vspace{0.5cm}

We used the same censoring mechanism and also the same hazards model as under the first setting. 
For this scenario, we considered sample sizes 3200 and 6400 when $\rho=0.3$, and sample sizes 1600 and 3200 when $\rho=0.5$.
Results, again based on 2000 runs for each configuration, are shown in Table 3. 
For the case $(n=3200,\rho=0.3)$ the coverage probability is a bit  too high  at $t=3$. In the other settings the estimator is unbiased and coverage is satisfactory.
The results concerning the constant effect estimator, $\hat\beta_X$, are reported in the second half of Table 4, and from there it is seen that the estimator is unbiased and that the variability is well estimated leading to satisfactory coverage probabilities. We also see that 2SLS estimator of Tchetgen et al. (2015) seems to be unbiased in this setting although there is no theoretical underpinning of this. To look further into this and to stress that the 2SLS  estimation
relies on a correct specification of
a model for the exposure $X$ given the instrument
$G$ we ran a final study as follows. The instrument $G$ was taken to be normally distributed with mean 2 and variance  $1.5^2$. The unobserved $U$ was taken to be $1.5Z^2$ with $Z$ generated as normal with mean 1 and variance $0.25^2$. The exposure $X$ was binary with 
$$
P(X=1|G,U)=\mbox{expit}\{-1+0.2G+0.5G^2+U-E(U)\}.
$$
In this way the correlation between $X$ and $G$ was approximately 0.56.
We generated $\tilde T$ according to the  hazard model
$$
E\left\{d\tilde N(t)|T\geq t, X,G,U\right\}=0.05+0.4X+0.3U,
$$
and censored all at $t=2$ resulting in approximately 25\% censorings. We used sample size $1000$ and $2000$ with 1000 runs for each configuration.
We calculated the 2SLS estimator in two ways using different  first stage models; we denote the 2SLS  estimator based on regressing $X$ on $G$ (despite that $X$ is binary) in the first stage by $\check\beta_{1X}$ and the 2SLS  estimator based on a first stage logistic regression model using $G$  as explanatory variable
by $\check\beta_{2X}$. We stress that the estimator suggested in this paper, $\hat \beta_X$, is not based on any modelling of $X$ given $G$ in contrast to the 2SLS estimator. Results are given in Table 5 where it is seen that the estimator $\hat \beta_X$ is unbiased while the two versions of the 2SLS estimator are both  biased.

\begin{table}
\caption{
Binary exposure case. 
Bias of $\hat B_{X}(t)$, average estimated standard error, sd($\hat B_{X}(t)$), empirical standard error, see($\hat B_{X}(t)$)), and coverage probability of 95\% pointwise confidence intervals CP($\hat B_{X}(t)$)) based on the instrumental variables estimator, in function of sample size $n$ and at different strengths $\rho$ (correlation) of the instrumental variable. Bias of $\tilde B_{X}(t)$ is the bias of the naive Aalen estimator.
 \bigskip}

{\small
\begin{center}
\begin{tabular}{|l|cccc|cccc|}
\hline
 &&\multicolumn{3}{c|}{ $\rho=0.3$}&&\multicolumn{3}{c|}{ $\rho=0.5$} \\
&n & $t=1$ & $t=2$ & $t=3$ & n& $t=1$ & $t=2$ & $t=3$\\

\hline
Bias $\hat B_{X}(t)$ &3200&0.000   &   0.001     &  -0.017&1600 &       -0.000&-0.005   &  -0.022     \\
sd ($\hat B_{X}(t)$)&&0.109 &0.194& 0.316& &0.102&0.183 &0.306\\
see ($\hat B_{X}(t)$)&&0.109 &0.194& 0.331& &0.102&0.183 &0.302\\
95\% CP($\hat B_{X}(t)$)&&95.3&     95.4&      96.6   &&   95.7 &95.6&     96.1\\
Bias $\tilde B_{X}(t)$ &&-0.082   &   -0.164     &  -0.248& &      -0.085&-0.167   &  - 0.249      \\
Bias $\hat B_{X}(t)$& 6400&-0.000   &   -0.006     &  -0.015&  3200 &    0.001&0.001   &   -0.005      \\
sd ($\hat B_{X}(t)$)&&0.077 &0.137& 0.221& &0.071&0.128 &0.202\\
see ($\hat B_{X}(t)$)&&0.077 &0.135& 0.216&& 0.072&0.128 &0.207\\
95\% CP($\hat B_{X}(t)$)&&95.1&     94.6&      95.2   &&   95.1 &95.2&     95.9 \\
Bias $\tilde B_{X}(t)$ &&-0.082  &   -0.167     &  -0.250&&       -0.083&-0.168  &   -0.253       \\
\hline
\end{tabular}
\end{center}
}
\end{table}
\bigskip
\bigskip

\nothere{
\begin{table}
\caption{Continuous  exposure case. 
Bias of $\hat B_{X}(t)$, average estimated standard error, sd($\hat B_{X}(t)$), empirical standard error, see($\hat B_{X}(t)$), and coverage probability of 95\% pointwise confidence intervals CP($\hat B_{X}(t)$)) based on the instrumental variables estimator, in function of sample size $n$ and at different strengths $\rho$ (correlation) of the instrumental variable. \bigskip}
{\small
\centering
\begin{tabular}{|ll|cccc|cccc|}
\hline
$n$& &\multicolumn{4}{c|}{ $\rho=-0.3$}&\multicolumn{4}{c|}{ $\rho=-0.5$} \\
& & $t=1$ & $t=2$ & $t=3$& $t=4$ & $t=1$ & $t=2$ & $t=3$& $t=4$ \\

\hline
&Binary $X$   & & &  & &&  & &\\
400&Bias $\hat B_{X}(t)$ &-0.004&-0.005&0.008&-0.026&0.001&0.001&0.001&-0.002\\
&sd ($\hat B_{X}(t)$)&0.152 &0.234& 0.310& 0.406&0.091 &0.137& 0.182& 0.225\\
&see ($\hat B_{X}(t)$)&0.151 &0.232& 0.310& 0.409&0.089 &0.135& 0.177& 0.221\\
&95\% CP($\hat B_{X}(t)$)&96.2 &97.3& 96.9& 97.0&95.1 &95.7& 95.5& 95.2\\
800&Bias $\hat B_{X}(t)$ &-0.005   &   -0.004     &  -0.012&       -0.018&0.001   &   0.000     &  -0.003&       -0.004  \\
&sd ($\hat B_{X}(t)$)&0.107 &0.158& 0.212& 0.263&0.065 &0.097& 0.128& 0.166\\
&see ($\hat B_{X}(t)$)&0.104 &0.159& 0.209& 0.262&0.063 &0.095& 0.124& 0.154\\
&95\% CP($\hat B_{X}(t)$)&95.3&     96.0&      96.0   &   96.4 &94.9&     94.5&      94.4   &   95.4 \\
1600&Bias $\hat B_{X}(t)$ &0.001   &   0.002     &  0.001&       -0.002&-0.000   &   0.001     &  0.000&       0.002  \\
&sd ($\hat B_{X}(t)$)&0.076 &0.115& 0.150& 0.187&0.040 &0.066& 0.085& 0.108\\
&see ($\hat B_{X}(t)$)&0.073 &0.111& 0.147& 0.183&0.044 &0.067& 0.088& 0.109\\
&95\% CP($\hat B_{X}(t)$)&94.8&     95.1&      95.7   &   95.6 &95.2&     95.4&      95.7   &   95.1 \\
&Continuous  $X$   & & &  & &&  & &\\
800&Bias $\hat B_{X}(t)$ &-0.002&-0.007&0.003&-0.020&0.000&0.001&0.000&-0.007\\
&sd ($\hat B_{X}(t)$)&0.090 &0.147& 0.227& 0.321&0.049 &0.081& 0.126& 0.174\\
&see ($\hat B_{X}(t)$)&0.092 &0.149& 0.240& 0.317&0.049 &0.081& 0.118& 0.218\\
&95\% CP($\hat B_{X}(t)$)&96.8 &96.4& 96.3& 95.7&94.9 &95.3& 93.6& 93.9\\
1600&Bias $\hat B_{X}(t)$ &-0.001   &   -0.002     &  0.000&       -0.009&0.000  &   0.000     &  -0.001&       -0.005  \\
&sd ($\hat B_{X}(t)$)&0.066 &0.105& 0.151& 0.207&0.035 &0.059& 0.086& 0.124\\
&see ($\hat B_{X}(t)$)&0.0651 &0.104& 0.146& 0.204&0.035 &0.058& 0.085& 0.117\\
&95\% CP($\hat B_{X}(t)$)&95.3&     96.0&      96.0   &   96.4 &95.8&     94.4&      95.0   &   93.8 \\
3200&Bias $\hat B_{X}(t)$ &0.000   &   0.000     & - 0.003&       -0.002&-0.000   &   0.001     &  0.001&       0.003  \\
&sd ($\hat B_{X}(t)$)&0.045 &0.074& 0.104& 0.139&0.025 &0.041& 0.061& 0.089\\
&see ($\hat B_{X}(t)$)&0.046 &0.073& 0.102& 0.141&0.025 &0.041& 0.061& 0.085\\
&95\% CP($\hat B_{X}(t)$)&95.4&     94.9&      95.3   &   96.0 &94.4&     95.7&      95.1   &   94.6 \\

\hline
\end{tabular}
}
\end{table}
\bigskip
\bigskip
}

\begin{table}
\caption{Summary of simulations concerning the constant parameter estimator $\hat \beta_X$. Binary and continuous exposure case. 
Bias of $\hat \beta_{X}$, average estimated standard error, sd($\hat \beta_{X}$), empirical standard error, see($\hat \beta_{X}$), and coverage probability of 95\% pointwise confidence intervals CP($\hat \beta_{X}$)) based on the instrumental variables estimator, in function of sample size $n$ and at different strengths $\rho$ (correlation) of the instrumental variable.
Results for the 2SLS estimator $\check\beta_X$ of Tchetgen et al. (2015) are also given. \bigskip}

{\small
\begin{center}
\begin{tabular}{|lcccc|}
\hline
Continuous $X$ &\multicolumn{4}{c|}{ $(n,\rho)$} \\
& (1600,0.3)&(3200,0.3)&(800,0.5)&(1600,0.5)\\
Bias $\hat \beta_X$ &-0.002 &-0.004& -0.003&0.001 \\
sd ($\hat \beta_X$)&0.107 &0.074& 0.082& 0.057\\
see ($\hat \beta_X$)&0.113 &0.073& 0.084& 0.057\\
95\% CP($\hat \beta_X$)&97.2 &95.5 & 96.1 & 95.5\\
Bias $\check \beta_X$ &0.003 &-0.001&-0.003&0.001 \\
sd ($\check\beta_X$)&0.098 &0.068& 0.075& 0.053\\
\hline
Binary  $X$ &\multicolumn{4}{c|}{ $(n,\rho)$} \\
& (3200,0.3)&(6400,0.3)&(1600,0.5)&(3200,0.5)\\
Bias $\hat \beta_X$ &-0.002 &-0.004& -0.003& -0.000\\
sd ($\hat \beta_X$)&0.085 &0.061& 0.082& 0.056\\
see ($\hat \beta_X$)&0.088 &0.062& 0.081& 0.057\\
95\% CP($\hat \beta_X$)&96.2 &95.4 & 95.5 & 95.4\\
Bias $\check \beta_X$ &0.001 &-0.001&-0.001&-0.002 \\
sd ($\check\beta_X$)&0.072 &0.050& 0.068& 0.048\\
\hline
\end{tabular}
\end{center}
}
\end{table}
\bigskip
\bigskip

\begin{table}
\caption{Summary of simulations concerning the constant parameter estimator $\hat \beta_X$ and two versions of the 2SLS estimator of Tchetgen et al. (2015). Binary exposure  and  continuous instrument.
Mean of $\hat \beta_{X}$, average estimated standard error, sd($\hat \beta_{X}$), 
in function of sample size $n$.
Results for two versions (see text for details)  of  2SLS estimator $\check\beta_{1X}$ and $\check\beta_{2X}$ of Tchetgen et al. (2015) are also given. \bigskip}

{\small
\begin{center}
\begin{tabular}{ccccccc}
\hline

n& mean $\hat \beta_X$ &sd ($\hat \beta_X$)&mean $\check \beta_{1X}$ &sd ($\check \beta_{1X}$) &mean $\check \beta_{2X}$ &sd ($\check \beta_{2X}$) \\
1000& -0.002&0.117&0.069&0.117&0.039&0.100\\
2000& -0.002&0.079&0.067&0.079&0.038&0.068\\
\hline
\end{tabular}
\end{center}
}
\end{table}
\bigskip
\bigskip


\begin{figure}[ht]
\label{HRS-causal}
\begin{center}
\includegraphics[width=12cm, height=8cm]{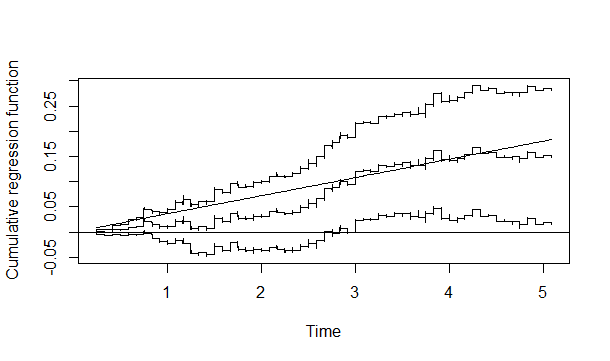}
 \caption{HRS-study. 
 Estimated causal effect of diabetes, $\hat B_X(t)$,   along with 95\% pointwise confidence bands. The straight line corresponds to the constant effects estimator (8).}
 \end{center}
 \end{figure}
 
 \subsection{Application to the HRS on causal association between diabetes and mortality}
We illustrate the proposed method using data from the Health and Retirement Study (HRS), a cohort initiated in 1992. The same data was used by Tchetgen Tchetgen et al. (2015)  (TT) to investigate the causal association between diabetes and mortality.
The HRS consists of persons ages 50 years or older and their spouses. There are genotype data for 12123 participants, but, like TT, we restrict our analyses to the 8446 non-Hispanic white persons with valid self-reported diabetes status at baseline. 
The average follow-up time was 4.10 years with a total of 644 deaths over 34055 person-years. We used an externally validated genetic risk score predictor of type 2 diabetes as IV. The risk score is based on 39 SNPs that were strongly associated with the  diabetes status, Likelihood ratio test chi-square statistic equal to 176.75 with 39 degrees of freedom, p-value < $10^{-6}$. Like TT we used as observed confounders ($L$) age, sex and the top 4 genomewide principal components to account for possible population stratification. The 2SLS control function approach used in TT is only valid if the instrument is binary unless one makes a further  linearity assumption concerning  a conditional mean of the un-observed confounder(s), specifically they assume that
$E\{\Omega(t,U)| G,X\}$ is linear in $G$. This  assumption is un-testable based on the observed data. The method we suggest in this paper is not restricted to only binary instruments. As a matter of fact no restrictions are put on neither the exposure nor the instrument. They can be binary as well as continuous. Also, the approach taken in TT assumes a time-constant exposure effect whereas the approach suggested in this paper allows the exposure effect to vary with 
time, and we may test whether a time-constant seems reasonable. The analysis used here thus generalizes that of TT in several aspects. Figure 2 shows the estimated causal effect of diabetes status on mortality, $\hat B_X(t)$,   along with 95\% pointwise confidence bands. The straight line corresponds to the constant effects estimator (8). From Figure 2 it seems reasonable to assume a time-constant exposure effect, which we can formally test using the statistic (9). This procedure gives a p-value of 0.61 thus giving no evidence against the time-constant  exposure effect model. 
The estimate of  the time-constant exposure effect is  $\hat\beta_X=0.036$ with estimated standard error   0.0142 corresponding to the 95\% confidence interval (0.008,0.064). So there seems to be a causal association between diabetes status and all cause mortality corresponding to  an average of 3.6 additional deaths occurring  for each year of follow-up in each 100 persons with diabetes alive at the start of the year, compared with each 100 diabetes-free  persons alive at the start of the year, conditional on age and sex.
 This estimated effect is less than half of that obtained by TT suggesting that the  linearity assumption used in TT  may not hold.

\subsection{Application to the HIP trial on effectiveness of screening  on breast cancer mortality}
The Health Insurance Plan (HIP) of Greater New York was a randomized trial of breast cancer screening that began in 1963. The purpose was to see whether 
screening has any effect on breast cancer mortality.
%
About 60000 women aged 40-60 were randomized into two approximately equally sized groups. Study women were offered the screening examinations consisting of clinical examination, usually by a surgeon, and a mammography. Further three annual examinations were offered in this group. Control women continued to receive their usual medical care. About 35\% of the women that were offered screening refused to participate  (non-compliers), see Table 5.
There were  large differences between the study women who participated and those who refused (Shapiro, 1977) and therefore the  results from the "as treated" analysis may be doubtful  due to unobserved confounding.  
\begin{table}[H]
\caption{HIP-study.  \bigskip}

{\small
\begin{center}
\begin{tabular}{|cc|ccc|}
\hline
& Control & \multicolumn{3}{c|}{Screening group} \\
& Group & &&\\
& & All &Compl. & Non-compl.  \\
\hline
n&30565&30130&20146&9984\\
\hline
\end{tabular}
\end{center}
}
\end{table}
\bigskip

\noindent
The same data were analysed by Joffe (2001) and as he did, we will also focus on the first 10 years of follow-up. 
Since screening ended after  three years, Joffe argued that focussing on the first 10 years of follow up will reduce attenuation of the effects of screening in the later
periods in which  treatment was the same both groups.
We can look into the possibility of a time-varying effect in a more formal way as our estimator $\hat B_X(t)$ captures this directly.
To begin with we performed a  Cox-regression intention to treat analysis showing that there is reduced mortality from breast cancer in the screening group (p=0.01). We also applied the  Aalen additive hazards intention to treat analysis.
Figure 2 shows the estimated cumulative regression coefficient along with 95\% confidence intervals indicating a time-varying effect of the screening; there seems to be a beneficial effect in the first 6 years or so, and no effect thereafter. The supremum test of an overall effect of screening is significant (p=0.005).
\begin{figure}
\label{Aalen-ITT}
\begin{center}
\includegraphics[width=12cm, height=8cm]{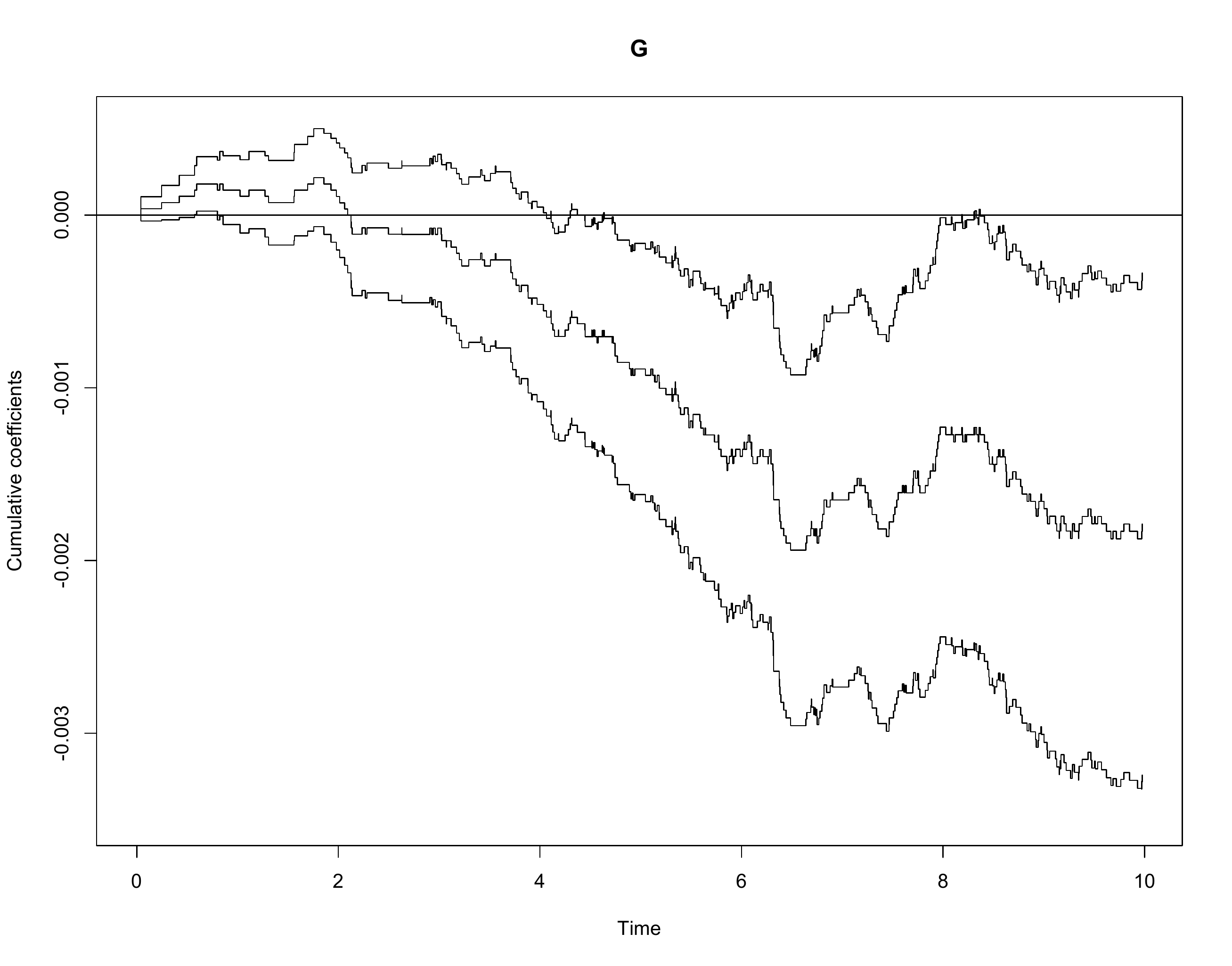}
 \caption{HIP-study. 
Aalen additive hazards  intention to treat analysis. Estimated cumulative regression coefficient   along with 95\% pointwise confidence bands .}
 \end{center}
 \end{figure}
 We will now apply our suggested method to estimate the causal effect of screening using the randomisation variable as instrument. 
 In our notation, the randomization variable is called $G$ and the treatment, screening, is called $X$.
Before proceeding, it is important to notice that there is a competing risk issue in these data. In the first 10 years of follow-up there are 4221 deaths but only 340 were deemed due to breast cancer. 
 The $i$th counting process in our estimator (7) is now the counting process that jumps at time point $t$ if the $i$th women at that  point in time dies from breast cancer.
We show in a separate report to be communicated elsewhere that  $\hat B_X(t)$ 
contrasts  the cumulative 
 breast cancer death specific hazards 
among the treated
between scenarios with versus without screening
 under the assumption that 
 the cause specific hazard of death due to other causes than breast cancer for the screened women would have been the same at all times had they not been screened.
 To test this assumption one may use the  test process
$$
H_n(t)=n^{-1/2}\sum_i\int_0^t(G_i-\overline{G})e^{\hat B_X(s-)X_i}dN_{2i}(s),
$$
where $N_{2i}(t)$ is the $i$th counting process counting  non-breast cancer death. Under the null of no causal effect of screening on the  non-breast cancer death hazards, this process is a zero-mean process. One may further show that 
$$
H_n(t)=n^{-1/2}\sum_i\epsilon_i^H(t)+o_p(1),
$$
where $\epsilon_i^H(t)$ are independent identically distributed zero-mean processes. 
Specifically,
$$
\epsilon_i^H(t)=\{G_i-E(G_i)\}\left \{\int_0^te^{ B_X(s)X_i}dN_{2i}(s)-\zeta_1(t)\right\}+\epsilon_i^B(t,\theta)\zeta_2(t)-\int_0^t\zeta_2(s)d\epsilon_i^B(s,\theta)
$$
considering here the case without covariates so that $\theta=E(G_i)$. In the previous display, $\zeta_1(t)$ and $\zeta_2(t)$ are the limits in probability of 
\begin{align*}
&n^{-1}\sum_i\int_0^te^{ B_X(s)X_i}dN_{2i}(s)\quad \mbox{and}\\
&n^{-1}\sum_i\int_0^t\{G_i-E(G_i)\}e^{ B_X(s)X_i}X_idN_{2i}(s),
\end{align*}
 respectively. This representation can be used to resample from the limit distribution of 
$H_n(t)$ under the null. Further, a formal test based on for instance  $\sup_{t\leq 10}|H_n(t)|$ may be performed and whether or not it is significant can also be 
\begin{figure}
\begin{center}
\includegraphics[width=12cm, height=8cm]{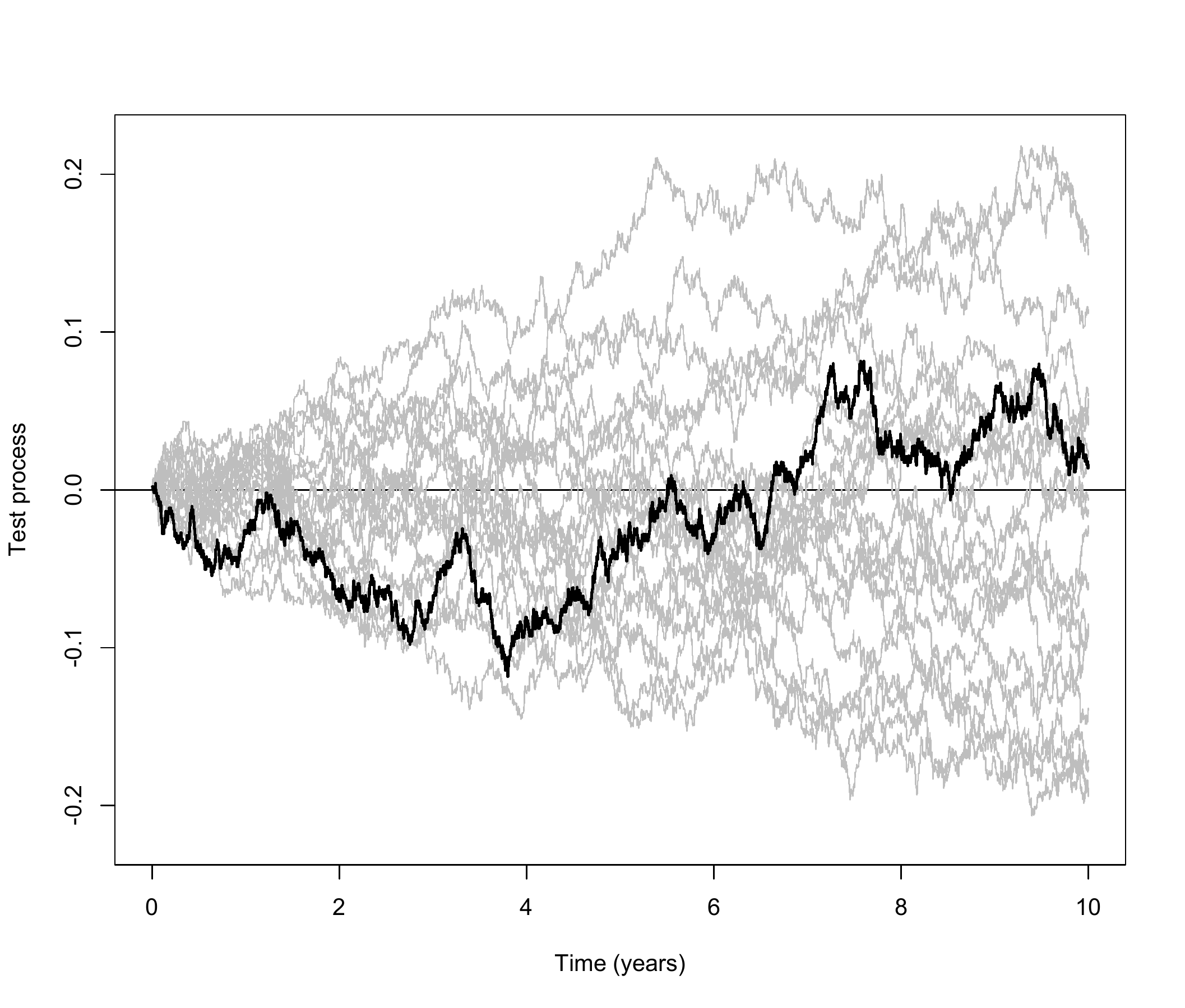}
 \caption{HIP-study. Investigation of 
 whether the cause specific hazard of death due to other causes than breast cancer for the screened patients would have been the same at all times had they not been screened.
Test process $H_n(t)$ along with 20 resampled processes from its limit distribution under the null.}
 \end{center}
 \end{figure}
based on resampling from the limit distribution under the null. Figure 3 shows the test process $H_n(t)$ along with 20 resampled processes from its limit distribution under the null, and it is seen that the test process does not seem to deviate in any respect. The supremum test based on 1000 resamples results in a p-value of 0.63. Based on this, we proceed to calculate the estimator $\hat B_X(t)$. This estimate along with 95\% confidence bands (pointwise) are given in Figure 4 that also shows the intention to treat estimate (broken curve). The causal effect of the screening appears to be slightly more pronounced than what is seen  from the intention to treat estimator and again it is seen that there seems to be a time-varying effect with screening being beneficial in a period of approximately 6 years. The supremum test 
$\sup_{t\leq 10}|\hat B_X(t)|$ is significant (p=0.02).

\begin{figure}
\begin{center}
\includegraphics[width=12cm, height=10cm]{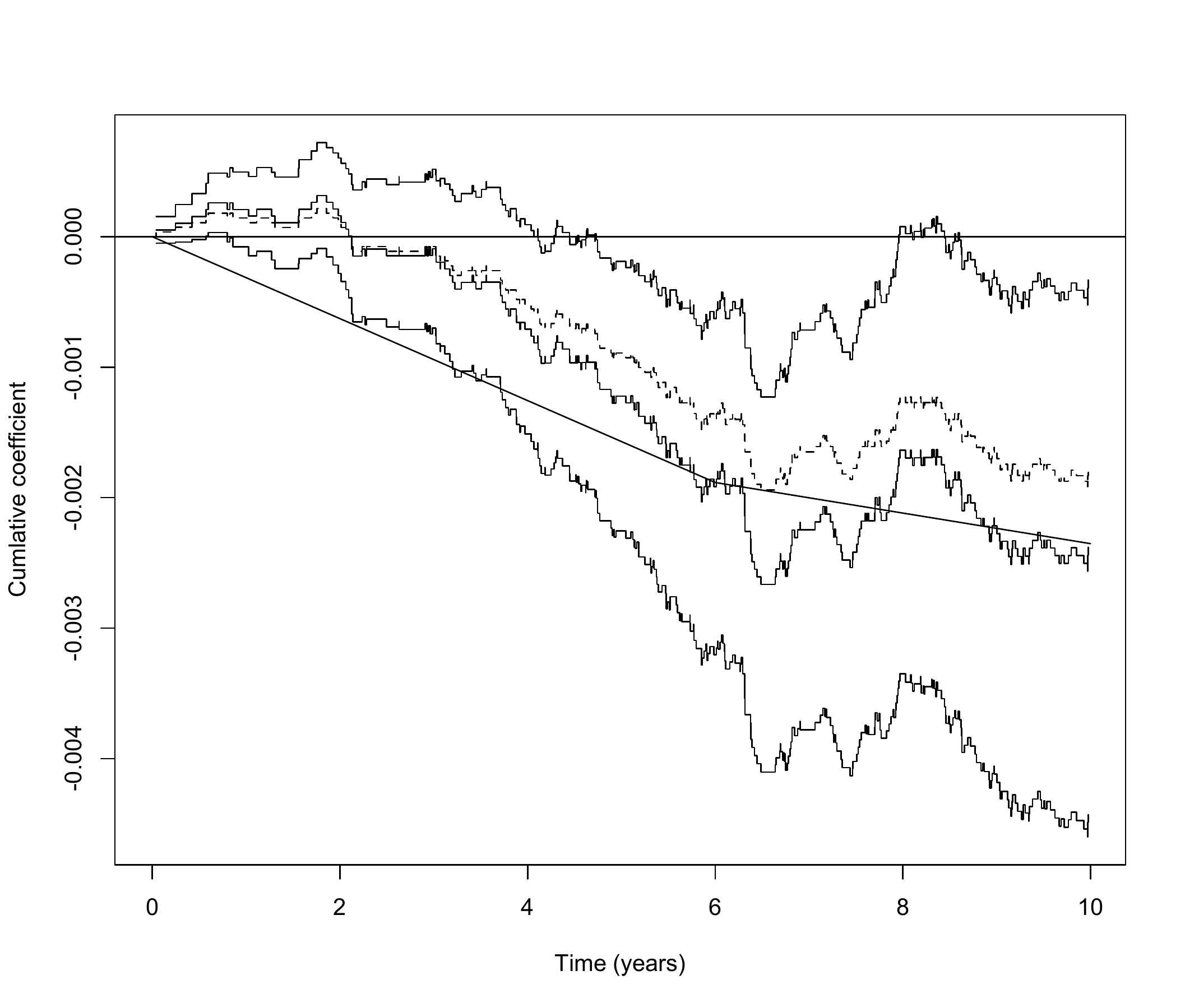}
 \caption{HIP-study. 
 Estimated causal effect of screening, $\hat B_X(t)$   along with 95\% pointwise confidence bands (solid curves) and the intention to treat estimate (broken curve). 
 A two-parameter piecewise constant estimator
 $B_X^{\dag}(t)$, see \eqref{Simpel_est},  is also shown.}
 \end{center}
 \end{figure}
 
 Using our approach it is now possible to study the time-dynamics even further. For illustrative purposes, let us assume that it had been hypothesized that if there were  an effect of screening it would only last for a few years (as screening stopped after 3 years), and let us say it corresponds to roughly six years of follow up.
We could then attempt the  simpler model
\begin{equation}
\label{SimpelM}
\beta_X(t)=\beta_0I(t<\xi)+\beta_1I(t\ge \xi).
\end{equation}
with $\xi=6$ years.
The two parameters $\beta_0$ and $\beta_1$ are estimated by
\begin{equation*}
\hat \beta_0=\int_0^{\xi} w_0(t)d\hat B_{X}(t)\quad \hat \beta_1=\int_{\xi}^{\tau} w_1(t)d\hat B_{X}(t)
\end{equation*}
with $w_0(t)=\tilde w(t)/\int_0^{\xi}\tilde w(s)\, ds$, $w_1(t)=\tilde w(t)/\int_{\xi}^{\tau}\tilde w(s)\, ds$,
$\tilde w(t)=R_{\mbf{\cdot}}(t)=\sum_iR_i(t)$. The estimate of $B_{X}(t)$ under this simplified model is then given by
\begin{equation}
\label{Simpel_est}
 B_{X}^{\dag}(t)=\hat \beta_0tI(t<\xi)+\hat\beta_0\xi I(t\ge \xi)+ \hat \beta_1(t-\xi)I(t\ge \xi),
\end{equation}
The constant effects parameters are estimated
 to $\hat\beta_0=-0.00031$  (SE 0.00011) and  $\hat\beta_1=-0.00012$ (SE 0.00020), indicating a significant effect of the screening only in the first  6 years. The estimator $
 B_{X}^{\dag}(t)$ is shown in Figure 4.
To test whether the simplified model, that is assuming a constant effect of treatment with a change in the effect after 6 years, gives a reasonable description of the data we consider the test process $TST(t)=n^{1/2}\{\hat B_{X}(t)- B_{X}^{\dag}(t)\}$ which, under the null, can be written as
\begin{align*}
n^{1/2}\{\hat B_{X}(t)- B_{X}^{\dag}(t)\}=&n^{1/2}\{\hat B_{X}(t)- B_{X}(t)\} -n^{1/2}(\hat \beta_0-\beta_0)tI(t<\xi)-\\
&n^{1/2}(\hat \beta_0-\beta_0)\xi I(t\ge\xi)-
n^{1/2}(\hat \beta_1-\beta_1)(t-\xi)I(t\ge \xi).
\end{align*}
Using the iid representation of $W_n(t)$ we can resample from the limit distribution, under the null, of $TST(t)$; such 20 randomly picked processes are shown in Figure 5 along with the observed test process $TST(t)$.
We may use the supremum test statistic 
$
TST=\sup_{t\le 10}|TST(t)|
$
  to investigate  
whether the test process is deviating.
%
To see whether the observed $TST$ is extreme we sampled 1000 draws from the limit distribution as outlined in Section 3; this gave a p-value of 0.56 suggesting that the constant effects model with a change in the effect after 6  years gives a reasonable fit to the data. 
\begin{figure}
\begin{center}
\includegraphics[width=12cm, height=8cm]{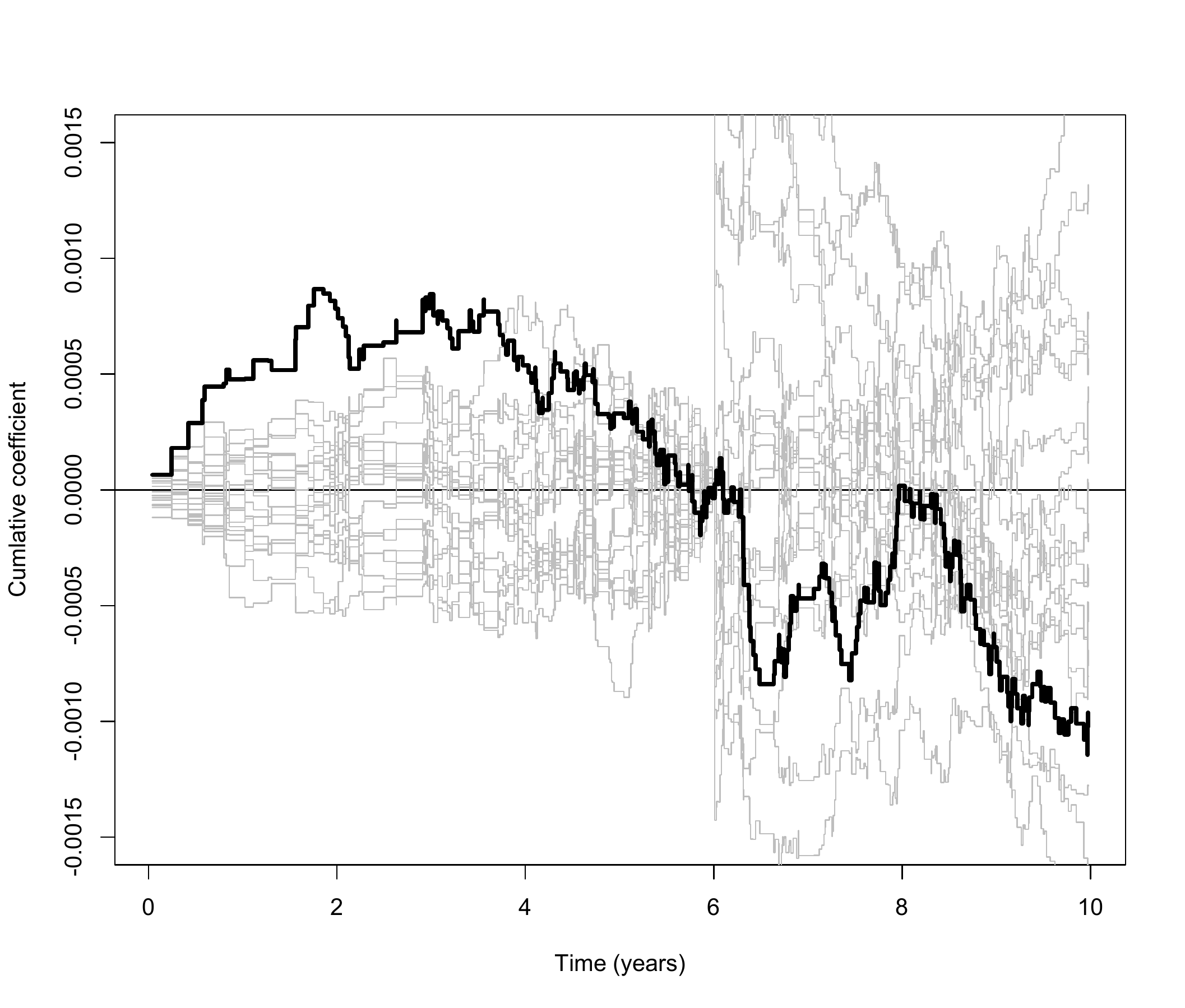}
 \caption{HIP-study. 
 Observed goodness-of-fit test process $TST(t)$ (thick curve)   along with  20 resampled processes under the null.}
 \end{center}
 \end{figure}
However, it also seen from Fig. 4 and Fig. 5 that the two parameter constant effects model is perhaps not giving a fully satisfactory fit in the first period of follow-up (two years or so). Actually, if instead one uses the test statistic 
$TST=\sup_{t\le 6}|TST(t)|$ 
then one gets a p-value of 0.06 giving some indication of a non-satisfactory fit in the initial phase of the follow up period. One could consider extending the two parameter constant effects model with an additional parameter allowing for a separate effect in the initial phase of two years or so. The cutpoints chosen here were used for illustrative purposes only, in practice they should have been specified ahead of performing the analysis.

\nothere{
\subsection{Application to the Illinois Reemployment Bonus Experiment}

The Illinois Reemployment Bonus Experiment study (see Bijwaard and Ridder (xx), and references therein) aimed to investigate whether bonuses paid to Unemployment Insurance (UI) benificiaries or their employers reduce the time spent in unemployment relative to a randomly selected control group. WHAT IS CONTROL? In the experiment, newly unemployed claimants were randomly divided into three groups: a Claimant Bonus Group, a Employer Bonus Group and, a control group. The members of both treatment groups were instructed that they (Claimant Bonus Group) or their employer (Employer Bonus Group) would qualify for a cash bonus  of \$500 if they found a job (of at least 30 hours) within 11 weeks and held that job for at least four months. Those randomized to the two treatment groups had the possibility to refuse participation in the experiment. The response variable is insured weeks of unemployment. WHAT DO YOU MEAN BY `INSURED'? Since UI benefits end after 26 weeks all unemployment durations 
are censored at 26 weeks.

\begin{table}
\caption{Average Unemployment Durations (AUD) when ignoring censoring. \bigskip}

{\small
\begin{center}
\begin{tabular}{|cc|ccc|ccc|}
\hline
& Control & \multicolumn{3}{c|}{Claimant group} &\multicolumn{3}{c|}{Employer group}\\
& Group & &&&&&\\
& & All &Compl. & Non-compl. & All &Compl. & Non-compl. \\
\hline
AUD &18.33 &16.96&16.74&18.18&17.65&17.62&17.72\\
n&3952&4186&3527&659&3963&2586&1377\\
\hline
\end{tabular}
\end{center}
}
\end{table}
\bigskip

\vspace{0.5cm}
\centerline{Table 4 about here}
\vspace{0.5cm}

\noindent

Table 4 gives the average unemployment durations ignoring the censoring. It shows that about 15\% of Claimant group and 35\% om the Employer group declined participation (for unknown reasons). Because the effect of participation may well be confounded (Bijwaard and Ridder, xx), we will here use an IV analysis with randomisation as the instrument. 
We will first focus on the comparison between the Claimant bonus group and the control group, and return to the Employer group later. We  first employ the estimator $\hat B_{\mbox{Cl}}(t)$ given in (\ref{Est})  where the treatment effect is allowed to vary freely with time. Next, because there may  be a time-changing treatment effect with a change in the effect after 10 weeks (in line with the end of the eligibility period of the bonuses) (Bijwaard and Ridder, xx)), we consider  the simpler model where 
\begin{equation}
\label{SimpelM}
\beta_{\mbox{Cl}}(t)=\beta_0I(t<\xi)+\beta_1I(t\ge \xi).
\end{equation}
The two parameters $\beta_0$ and $\beta_1$ are estimated by
\begin{equation*}
\hat \beta_0=\int_0^{\xi} w_0(t)d\hat B_{\mbox{Cl}}(t)\quad \hat \beta_1=\int_{\xi}^{\tau} w_1(t)d\hat B_{\mbox{Cl}}(t)
\end{equation*}
with $w_0(t)=\tilde w(t)/\int_0^{\xi}\tilde w(s)\, ds$, $w_1(t)=\tilde w(t)/\int_{\xi}^{\tau}\tilde w(s)\, ds$,
$\tilde w(t)=R_{\mbf{\cdot}}(t)=\sum_iR_i(t)$. The estimate of $B_{\mbox{Cl}}(t)$ under this simplified model is then given by
$$
\tilde B_{\mbox{Cl}}(t)=\hat \beta_0tI(t<\xi)+\hat\beta_0\xi I(t\ge \xi)+ \hat \beta_1(t-\xi)I(t\ge \xi),
$$
Figure 2 gives the estimator $\hat B_{\mbox{Cl}}(t)$ along with 95\% pointwise confidence bands and also the estimator  $\tilde B_{\mbox{Cl}}(t)$ using the null model given by \eqref{SimpelM}. There seems to a significant causal effect of the Claimant bonus reducing time spend in unemployment. TORBEN, IF WE ONLY LOOK AT SIGNIFICANCE, THEN THE ITT ANALYSIS WILL BE JUST AS GOOD. TO SHOW THE BENEFITS OF THE PROPOSED METHODS, WE SHOULD THEREFORE EMPHASISE THE INTERPRETATION OF THE MAGNITUDE OF THE EFFECT, RATHER THAN JUST THE SIGNIFICANCE. I WOULD FIND IT USEFUL TO SEE THE EFFECT IN THE FIGURES ON THE RELATIVE RISK SCALE FOR THAT PURPOSE, THAT IS, EXPONENTIATED.

\begin{figure}
\begin{center}
\includegraphics[width=12cm, height=10cm]{/Users/dkn606/Documents/Arbejde/papers/InstrumentalVariables/Clayman_vs_Control_const10.pdf}
 \caption{ Illinois Reemployment Bonus Experiment data. 
Estimated  causal effect of Claimant bonus, $\hat B_{\mbox{Cl}}(t)$,  along with 95\% confidence bands 
 and the estimate, $\tilde B_{\mbox{Cl}}(t)$, under the simplified model. }
 \end{center}
 \end{figure}

\noindent
To test whether the simplified model, that is assuming a constant effect of treatment with a change in the effect after 10 weeks, gives a reasonable description of the data we consider the test process $TST(t)=n^{1/2}\{\hat B_{\mbox{Cl}}(t)-\tilde B_{\mbox{Cl}}(t)\}$ which, under the null, can be written as
\begin{align*}
n^{1/2}\{\hat B_{\mbox{Cl}}(t)-\tilde B_{\mbox{Cl}}(t)\}=&n^{1/2}\{\hat B_{\mbox{Cl}}(t)- B_{\mbox{Cl}}(t)\} -n^{1/2}(\hat \beta_0-\beta_0)tI(t<\xi)-\\
&n^{1/2}(\hat \beta_0-\beta_0)\xi I(t\ge\xi)-
n^{1/2}(\hat \beta_1-\beta_1)(t-\xi)I(t\ge \xi).
\end{align*}
Using the iid representation of $W_n(t)$ we can resample from the limit distribution, under the null, of $TST(t)$; such 25 randomly picked processes are shown in Figure 3 along with the observed test process $TST(t)$ that does not seem to show a deviating behaviour.
To investigate  this more formally we can  use the test statistic 
$$
TST=\sup_{t\le \tau}|TST(t)|
$$
to test the goodness-of-fit of model \eqref{SimpelM}.
To see whether the observed $TST$ is extreme we sampled 1000 draws from the limit distribution as outlined in Section 3; this gave a p-value of 0.36 suggesting that the constant effects model with a change in the effect after 10 weeks gives a reasonable fit to the data. The constant effects parameters are estimated to $\hat\beta_0=0.0093$  (SE 0.0020) and  $\hat\beta_1=0.0016$ (SE 0.0016), indicating a significant effect of the Claimant bonus only in the first 10 weeks. We conclude that for those who participated in the Claimant Bonus Group, the risk of being unemployed at 10 weeks would have been $\exp(10\hat\beta_0)=1.1$ (95\% confidence interval 1.06 to 1.14) times higher had they not participated.

\begin{figure}
\begin{center}
\includegraphics[width=12cm, height=10cm]{/Users/dkn606/Documents/Arbejde/papers/InstrumentalVariables/Clayman_vs_Control_const10_GOF}
 \caption{ Illinois Reemployment Bonus Experiment data. 
Observed goodness-of-fit test process (thick curve)   along with  25 resampled processes under the null.}
 \end{center}
 \end{figure}

To investigate whether the other treatment (Employer bonus) has a similar effect we will end this application by comparing the two treatment groups under the assumption of no current treatment interaction. TORBEN, IN PRINCIPLE, THIS COMPARISON IS ONLY USEFUL IF WE MAKE THE NON CURRENT TREATMENT INTERACTION ASSUMPTION. THE REASON IS THAT THE DIFFERENT ESTIMATES DESCRIBE EFFECTS FOR DIFFERENT SUBGROUPS, NAMELY THOSE WHO PARTICIPATED IN EITHER OF THE TWO TREATMENTS. I HAVE THEREFORE ADDED THIS ASSUMPTION AT THE END OF THE PREVIOUS LINE. We hence also compared the Employer bonus group to the control group obtaining the estimate $\hat B_{\mbox{Em}}(t)$. We then calculated the difference $\hat B_{\mbox{Cl}}(t)-\hat B_{\mbox{Em}}(t)$. The uncertainty of this estimate can be calculated using the iid representation of both $\hat B_{\mbox{Cl}}(t)$ and $\hat B_{\mbox{Em}}(t)$ keeping track of which individuals belongs to which group when doing the subtraction of the iid representations. The estimate of 
$\hat B_{\mbox{Cl}}(t)-\hat B_{\mbox{Em}}(t)$ is given in Figure 4 along with 95\% pointwise confidence bands showing no evidence  of a significant different effect of the two bonus treatments. The 
constant effects parameters for the Employer bonus group are estimated to $\hat\beta_0=0.0092$  (SE 0.0027) and  $\hat\beta_1=0.0025$ (SE 0.0020).

\begin{figure}
\begin{center}
\includegraphics[width=12cm, height=10cm]{/Users/dkn606/Documents/Arbejde/papers/InstrumentalVariables/Claiman_vs_Empl}
 \caption{ Illinois Reemployment Bonus Experiment data. 
Estimated difference between  effect of Claimant bonus and Employer bonus, $\hat B_{\mbox{Cl}}(t)-\hat B_{\mbox{Em}}(t)$,  along with 95\% confidence bands .}
 \end{center}
 \end{figure}
 }

\nothere{
\subsection{Causal effect of Vitamin D on total mortality}

Vitamin D is a fat-soluble vitamin retained from the diet and dietary supplements and produced in sun-exposed skin. Vitamin D deficiency is common and has been linked with a number of common diseases 
 such as cardiovascular disease, diabetes, and cancer. Here we will be interested in evaluating the causal effect on overall survival.  Unfortunately, vitamin D status is also associated with several behavioral and environmental factors potentially giving rise to biased estimators when using standard statistical analyses. 
Recently, mutations in the filaggrin gene have been shown to be associated with a higher vitamin D status, supposedly through an increased UV sensitivity of keratinocytes (see Skaaby et al., 2013, and references therein). 
Filaggrin is an important component of the skin barrier function. Loss-of-function mutations in the filaggrin gene affect 8-10\% of Northern Europeans (Palmer et al., 2006). The two mutations, 2282del4 and R501X, are by far the most frequent in white Caucasians, but the R2447X mutation also occurs. 

To assess the potential causal effect of vitamin D status on overall survival, we used filaggrin genotype as a proxy measure for vitamin D status according to the principles of Mendelian randomisation. The data source is the  population based study Monica10, and vitamin D is measured as  the serum 25-OH-D  (nmol/l).
The Monica I study was conducted in 1982-1984 and included examinations of 3785 individuals of Danish origin recruited from the Danish Central Personal Register as a random sample of the population. The ten-year follow-up study, Monica10, included 2656 individuals between 40-71 years. It was carried out in 1993-1994, for further details see  Olsen et al. (2007) and Skaaby et al. (2013).
We considered the following variables: $G$ is an indicator of filaggrin mutation; 
$X$ is  Vitamin D  (nmol/l) centered around the mean (65  nmol/l) and divided by its standard deviation (27 nmol/l).
 \begin{figure}
\begin{center}
\includegraphics[width=12cm, height=10cm]{/Users/dkn606/Documents/Arbejde/papers/InstrumentalVariables/Fig2.pdf}
 \caption{Monica10 data. Estimated cumulative regression coefficients along with 95\% pointwise confidence bands for Vitamin D using the naive Aalen additive hazards model.}
 \end{center}
 \end{figure}
 Our analysis below is a simplification of a complex reality and must therefore be viewed as an illustration, for various reasons. We only have information on the 
 Vitamin D exposure  at one point in time, so that we could not accommodate its time-varying nature. Filaggrin deficiency is associated with an increased risk of ichtyosis vulgaris, atopic dermatitis, allergic rhinitis, and asthma (van den Oord and Sheikh, 2009), with asthma itself being linked with mortality; this may imply a violation of the IV assumptions. Finally, as in all Mendelian randomisation analyses, our analysis is restricted to individuals who were alive at study entry. This could induce survivorship bias if 
 the effect of $G$ on death prior to study entry were non-negligible (Bochud and Rousson, 2010; Boef et al., 2015). The fact that in our study, no association was detected between the instrument and measured confounders gives confidence that the IV may be independent of (unmeasured) confounders within survivors  (Table 1 in Skaaby et al., 2013). 
 \begin{figure}
\begin{center}
\includegraphics[width=12cm, height=10cm]{/Users/dkn606/Documents/Arbejde/papers/InstrumentalVariables/Fig3.pdf}
 \caption{Monica10 data. Estimated  causal effect of Vitamin D, $\hat B_X(t)$,  along with 95\% pointwise confidence bands (full lines) and the naive estimate from Figure 6 (dashed line). }
 \end{center}
 \end{figure}
The time-scale is time since investigation, and the number of individuals with complete information on the 
variables examined was 2571. The odds of having a normal Vitamin D value (>30 nmol/l) was estimated to be 3.13 (95\% CI: 1.28 to 7.76, P=0.01)
 times larger for those with the filaggrin mutation compared to those without. 
 The $F$-test statistic, when regressing Vitamin D on the instrument, is 7.34 (p=0.007), demonstrating that the instrument is weak.
 The result concerning the observational effect of vitamin D using the naive Aalen additive hazards analysis, i.e. using the following model, where we also include
  $L=$age (in years) at investigation,
$$
\lambda_T(t|X,L)=\phi_0(t)+\phi_X(t)X+\phi_L(t)L,
$$
is displayed in Figure 6 giving the estimated cumulative regression functions $\int_0^t \phi_X(s)\, ds$ along with 95\% pointwise confidence bands. 
Judging from Figure 6 there seems to be an effect of Vitamin D on overall survival, but note that this is under the assumption of no unmeasured confounders. This assumption is likely violated because, for instance, people with poor health conditions might be less likely exposed to sunlight and thus have lower Vitamin D levels. 
Figure 7 displays the naive effect of Vitamin D again and also the proposed estimator $\hat B_X(t)$ estimating the conditional causal effect pointing to a much more profound effect of Vitamin D. However, the associated pointwise 95\% confidence bands are very wide indicating that there may not be any effect of Vitamin D at all. 
  \begin{figure}
\begin{center}
\includegraphics[width=12cm, height=10cm]{/Users/dkn606/Documents/Arbejde/papers/InstrumentalVariables/Fig4.pdf}
 \caption{Monica10 data. Estimated  causal effect of Vitamin D. Estimate, $e^{\hat B_X(t)}$, of ratio of  survival probabilities, $\frac{P(\tilde T^0>t| X=1, G,A)}{P(\tilde T>t| X=1, G,A)}$,  along with 95\% pointwise confidence bands.}
 \end{center}
 \end{figure}
 The standard test of effect was performed using an additive hazards analysis including the filaggrin variable and age as explanatory variables. The test of no effect of the filaggrin variable gave a p-value of 0.12 thus pointing in the same direction. 
 In Figure 8 we have plotted $e^{\hat B_X(t)}$ along with 95\% pointwise confidence bands; $e^{\hat B_X(t)}$ estimates 
$P(\tilde T^0>t| X=1, G,A)/P(\tilde T>t| X=1, G,A)$. 
We see from Figure 8, for indivduals at a given age and with a given genotype and  one standard deviation above the  mean on  the vitamin D scale, that if their  Vitamin D status had been set to the mean  then this would lead to a roughly 20 percent lower survival probability after 7 years. However, as already noted, this result is not significant.
}

\nothere{
\begin{figure}
\begin{center}
\includegraphics[width=12cm, height=10cm]{/Users/stijnvansteelandt/DropBox/Articles/Torben-IV/Fig2.pdf}
 \caption{Monica10 data. Estimated cumulative regression coefficients along with 95\% confidence bands using the naive Aalen additive hazards model.}
 \end{center}
 \end{figure}

 \begin{figure}
\begin{center}
\includegraphics[width=12cm, height=10cm]{/Users/stijnvansteelandt/DropBox/Articles/Torben-IV/Fig3.pdf}
 \caption{Monica10 data. Estimated  causal effect of Vitamin D, $\hat B_X(t)$,  along with 95\% confidence bands (full lines) and the naive estimate from Figure 2 (dashed  line). }
 \end{center}
 \end{figure}
}

\nothere{
\textit{Stijn: Should we also add something about that in the naive analysis the time axis is also somewhat arbitrary as it is the time from the time where vit d is measured to death (or censoring) whereas in our analysis we mimick  a randomized trial changing the exposure level, and the time from that to death (or censoring) makes a lot more sense. 
TORBEN, I AM NOT SURE I GET YOUR POINT SINCE WE ALSO COUNT TIME SINCE STUDY ENTRY. I SEE 2 OPTIONS. EITHER WE STATE THAT WE ARE MAKING THE IMPLICIT ASSUMPTION THAT VITAMIN D DOES NOT AFFECT MORTALITY DURING TIME TO STUDY ENTRY. OR WE ADD AN ANALYSIS WHICH COUNTS TIME SINCE BIRTH, BUT THEN WE WILL NEED TO PARAMETERIZE HOW THE EFFECT CHANGES WITH TIME, AS THERE IS NO INFORMATION ABOUT THE EFFECTS AT AGE BELOW 40.}
}
\section{Concluding remarks}


In this article, we  proposed an instrumental variables estimator for the effect of an arbitrary exposure on an event time. In comparison with other instrumental variables estimators for event times, our proposed approach has the advantage that it can handle arbitrary (e.g., continuous) 
exposures, without the need for modelling the exposure distribution, and that
it naturally adjusts for censoring whenever censoring is independent of the event time, exposure and instrument, conditional on measured and unmeasured confounders. The independent censoring assumption is relatively weak as it allows for a dependence on unmeasured factors. This 
assumption can be relaxed via inverse probability of censoring weighting under a model for the dependence of   censoring on
the exposure and/or instrumental variable. 

Under the usual instrumental variable assumptions, listed in Section 1, the IV-estimator (8) provides a consistent estimator of the causal exposure effect as opposed to the naive estimator when there is unmeasured confounding. However, in the case of a weak instrument, the IV-estimator may have a large variance.
 It is therefore of interest to develop semi-parametric efficient estimators  (Tsiatis, 2006). Along the same lines, it is also of interest to consider estimators that are robust to some model deviations. For instance, consider the following two models 
\begin{equation}
\label{DR-eq1}
\lambda_{\tilde{T}^0}(t|L)-\lambda_{\tilde{T}^0}(t|L=0)=\psi^T(t)L,
\end{equation}
and 
\begin{equation}
\label{DR-eq2}
E\{h(t,G,L)|L\}=E\{h(t,G,L)|L;\theta\},
\end{equation}
where  $h$ is a user defined function such as $h(t,G,L)=G$; and $\psi(t)$ and $\theta$ are parameters indexing the two models. 
Consider then the estimating function
\begin{equation}
\label{DR}
d(t,L)\left\{h^{*}(t,G,L)-\overline{h^{*}}(t)\right\}e^{B_X(t)X}R(t)\left\{dN(t)-dB_X(t)X-\psi^T(t)Ldt\right\},
\end{equation}
where 
$$
h^{*}(t,G,L)=h(t,G,L)-E(h(t,G,L)|L;\theta),
$$
$$
\overline{h^{*}}(t)=\frac{E\{h^{*}(t,G,L)R(t)e^{B_X(t)X}\}}{E\{R(t)e^{B_X(t)X}\}},
$$
and 
$d(t,L)$ is an  arbitrary index function.
One may then show that \eqref{DR} has zero mean  if either model 
 \eqref{DR-eq1}  or  model   \eqref{DR-eq2}  hold; the solution to an estimating equation based on estimating function (\ref{DR}) therefore yields a double robust estimator. This estimator has the further advantage of being invariant to linear transformations of the exposure.
A detailed study of efficient and double robust estimators  will be communicated in a separate report. 




\section*{Acknowledgement}
Torben Martinussen's work is part of the Dynamical Systems Interdisciplinary Network, University of Copenhagen.
Stijn Vansteelandt was supported by IAP research network grant nr. P07/05 from the Belgian government (Belgian Science Policy).	
Eric Tchetgen Tchetgen is supported by NIH grant R01A I104459.


\section*{Appendix: Large sample properties}

Let $\mu(L;\theta)=E(G|L;\theta)$ be the conditional mean of the instrument given observed confounders $L$, which is function of an unknown finite-dimensional parameter $\theta$. In the case of no observed confounders $\mu(\theta)=\theta=E(G)$ and $\hat \theta=\overline{G}$. We assume that $n^{1/2}(\hat\theta-\theta)=n^{-1/2}\sum_i\epsilon^{\theta}_i+o_p(1)$, where the $\epsilon^{\theta}_i$'s are zero-mean iid variables. In the case of no observed confounders  we have $\epsilon^{\theta}_i=G_i-\theta$. Let $\theta_0$ denote the true value of $\theta$.
\bigskip
\medskip

\noindent
We write $\nrm{g} = \sup_{t \in [0,\tau]} |g(t)|$ and use the notation $\cV(g)$ to denote the total variation
of $g$ over the interval $[0,\tau]$.
Let $B^\circ(t)$ denote the true value of $B(t)$, and let $M^\circ = \nrm{B^\circ}<\infty$.\\

\noindent
Technical  conditions:
\begin{itemize}
\item[(i)] We assume that $X$ and $G$ are bounded, and denote the respective bounds by $X_{max}$ and $G_{max}$.
\item[(ii)] Define $a(s,h) = E[R(s)XG^c e^{h X}]$. We assume that
there exist $M > M^\circ$ and $\nu>0$ such that
$\inf_{s \in [0,\tau], h \in[-M,M]} a(s,h) \geq 1.01 \nu$. \\

\end{itemize}
The quantities $M^\circ$ and $M$ do not necessarily need to be known.

\noindent
{\bf Consistency}
\medskip

\noindent
Below we show that $\hat B_X(t,\theta_0)$ is uniformly consistent. In what follows we suppress $\theta_0$
 from the notation and write $B(t)$ instead
of $B_X(t)$.
The estimator is given by the recursion equation
\begin{equation}
\hB(t) = \int_0^t \frac{\sum_i G_i^c e^{\hB(s-)X_i} dN_i(s)}
{\sum_i R_i(s)X_i G_i^c e^{\hB(s-)X_i}}
\end{equation}
It appears difficult to prove directly that $\hB(t)$ is bounded.
Instead we will take a different approach. We will modify the estimator in a way that will force
it to be of bounded variation.
We will then prove that the modified version of the estimator is consistent. 
If $M$ is not known, the modified estimator is a theoretical construct that cannot actually be computed, but it will emerge that for large enough $n$ the modified estimator  is equal to the unmodified estimator.

We will use the Helly Selection Theorem in the following form.

\it{Helly Selection Theorem}\rm: Let $\{f_n\}$ be a sequence of functions on $[0,\tau]$
such that $\nrm{f_n} \leq A_1$ and $\cV(f) \leq A_2$, where $A_1$ and $A_2$ are
finite constants. Then

\hspace*{24pt} a. There exists a subsequence $\{f_{n_j}\}$ of $\{f_n\}$ which converges pointwise to some function $f$.

\hspace*{24pt} b. If $f$ is continuous, the convergence is uniform.



\medskip
\noindent
Then it follows that $\nrm{\hB - B^\circ} \cas 0$.
\medskip

\noindent
\it{Proof}\rm: For a function $H(t)$ on $[0,\tau]$, define
\begin{align}
\Upsilon_n(H,t) & = \int_0^t \frac{n^{-1} \sum_i G_i^c e^{H(s-)X_i} dN_i(s)} {A(s,H(s-))} \\
\Upsilon(H,t) & = \int_0^t \frac{c(s,H(s))}{a(s,H(s))} \, ds
\end{align}
where
\begin{align}
A(s,h) & = \frac{1}{n} \sum_{i=1}^n R_i(s) X_i G_i^c e^{h X_i} \\
c(s,h) & = E[R(s) G^c e^{hX} \lambda(s,L,G,X)]
\end{align}
with $\lambda(s,L,G,X) = (d/ds) E[N(s)|L,G,X]$, so that $E[R(s) G^c e^{hX} dN(s)] = c(s,h) ds$.
The estimator $\hB(t)$ is then the solution to $B(t) = \Upsilon_n(B,t)$.
Let $\xi(y) = \mbox{sgn}(y) \min(|y|,M)$.
We then define the modified estimator $\tB$ to be the solution to the equation $B(t) = \Upsilon_n(\xi(B),t)$.
Note that $\Upsilon(\xi(B^\circ),t)=\Upsilon(B^\circ,t)=B^\circ(t)$.

Define $q(s,h) = c(s,\xi(h))/a(s,\xi(h))$, so that
$$
\Upsilon(\xi(H),t) = \int_0^t q(s,H(s)) ds
$$
The function $q(s,h)$ satisfies $\sup_{s\in[0,\tau],h \in \real} |q(s,h)| \leq 2 G_{max} e^{M X_{max}} \lambda_{max}\nu^{-1}$,
where $\lambda_{max}$ is an upper bound on $\lambda(s,L,G,X)$ (which we assume exists). Moreover, $q(s,h)$ is Lipschitz
with respect to $h$ over $s \in [0,\tau]$ and $h \in \real$ with Lipschitz constant
$\kappa = 2 G_{max} e^{M X_{max}} \lambda_{max}\nu^{-1} (1+ X_{max} G_{max} e^{M X_{max}} \nu^{-1})$.
Accordingly, by classical differential equations theory (Hartman, 1973, Thm.\ 1.1; Coddington, 1989,
Sec.\ 5.8), $B^\circ$ is the \it{unique} \rm solution to the equation $B(t)=\Upsilon(\xi(B),t)$ subject to $B(0)=0$.

We note for later reference that for any two functions $B_1$ and $B_2$ we have
\begin{equation}
\nrm{\Upsilon(\xi(B_1)) - \Upsilon(\xi(B_2))} \leq \kappa \tau \nrm{B_1-B_2}
\label{blip}
\end{equation}
Now, by the functional central limit theorem as given in Andersen and Gill (1982),
\begin{equation}
\sup_{s \in [0,\tau], h\in[-M,M]} |A(s,h) - a(s,h)| \stackrel{a.s.}{\rightarrow} 0
\label{fclt}
\end{equation}
Accordingly, from the the assumption that $\inf_{s \in [0,\tau], h\in[-M,M]} a(s,h) \geq 1.01\nu$, we get the result that
$\inf_{s \in [0,\tau], h\in[-M,M]} A(s,h) \geq \nu$ for $n$ sufficiently large. We thus find that the jumps in $\tB(t)$
are bounded by $n^{-1} D$ with $D=2 G_{max} e^{M X_{max}} / \nu$, implying that $\nrm{\tB} \leq D$ and
$\cV(\tB) \leq D$. Let $\mathbb{B}^*$ denote the class of functions $B(t)$ with these two properties. Further,
let $\mathbb{H}$ denote the class of functions that are bounded by $\tilde{M} = \min(M,D)$ and have total variation less than
$D$. Since $|\xi(y)| \leq |y|$ and $\xi$ is Lipschitz(1), we find that $B \in \mathbb{B}^*$ implies that $\xi(B) \in \mathbb{H}$.

Next, define
\begin{equation}
\tUps_n(H,t) = \int_0^t \frac{n^{-1} \sum_i G_i^c e^{H(s-)X_i} dN_i(s)} {a(s,H(s-))}
\end{equation}
From (\ref{fclt}) it follows that
\begin{equation}
\sup_{s \in [0,\tau], H \in \mathbb{H}} |\Upsilon_n(H,s) - \tUps_n(H,s)| \cas 0
\end{equation}

For $U=(T,\delta,X,L,G)$, define
\begin{equation}
\psi_{H,t}(U) = \frac{\delta G^c e^{H(T-)X}}{a(T,H(T-))}
\end{equation}
We then have $\tUps_n(H,t) = \mathbb{P}_n \psi_{H,t}$.
We claim that the class of functions $\mathcal{F} = \{\psi_{H,t},$ $H \in \mathbb{H}, t \in [0,\tau]\}$ is Donsker.
This result is an immediate consequence of the following facts:

\hspace*{24pt} 1. Sums and products of bounded Donsker classes are also Donsker.

\hspace*{24pt} 2. For any finite $K$, the class of monotone functions mapping $[0,\tau]$ to $[-K,K]$ is Donsker
(Kosorok, 2008, Thm.\ 9.24).

\hspace*{24pt} 3. If $H$ is bounded and has bounded variation, then $H$ can be written as $H = H_1 - H_2$,
where $H_1$ and $H_2$ are monotone increasing functions with $\nrm{H_1} \leq \nrm{H} + \cV(H)$
and $\nrm{H_2} \leq \cV(H)$ (Jordan decomposition).
It follows that the class of functions $H$ with $\nrm{H} \leq C_1$ and $\cV(H) \leq C_2$ is Donsker.

\hspace*{24pt} 4. If $H \in \mathbb{H}$, then the function $g(t) = a(t,H(t-)) = E[R(t)XG^c e^{H(t-)X}]$ is
bounded and of bounded variation with $\nrm{g} \leq 2 G_{max} e^{\tilde{M} X_{max}}$ and 
$$\cV(g) \leq
2 X_{max} G_{max} e^{\tilde{M} X_{max}} (\cV(r)+\cV(H)),$$ where $r(s) = E[R(s)]$.

It follows that
\begin{equation}
\sup_{t \in [0,\tau], H \in \mathbb{H}} |\tUps_n(H,t) - \Upsilon(H,t)| \cas 0
\end{equation}
and therefore
\begin{equation}
\sup_{t \in [0,\tau], H \in \mathbb{H}} |\Upsilon_n(H,t) - \Upsilon(H,t)| \cas 0
\label{Ups}
\end{equation}

Now, by Helly's selection theorem, every subsequence of $\tB(t)$ has a further subsequence that converges to some limit.
Since the jumps $\tB(t)$ are bounded by $n^{-1} D$ and the number of jumps in the interval $[t_1,t_2]$ divided by $n$
converges uniformly to $E[N(t_2)] - E[N(t_1)] \leq C (t_2 - t_1)$ for some constant $C$, it follows that the limit
of the sub-subsequence is continuous, and therefore (by the second part of Helly's theorem) the convergence
of the sub-subsequence is uniform.
Going further, the fact that $\tB = \Upsilon_n(\xi(\tB))$ in combination with (\ref{blip}) and (\ref{Ups}) implies that the limit
$B$ of the sub-subsequence satisfies $B = \Upsilon(\xi(B))$. But we said before that $B^\circ$ is the unique continuous
solution to this equation. We thus find that every subsequence of $\tB$ has
a further subsequence that converges uniformly to $B^\circ$. Consequently, $\tB$ itself converges uniformly to $B^\circ$.
Since $B^\circ \leq M^\circ$ and $\nrm{\tB - B^\circ} \cas 0$ (as just stated), for sufficiently large $n$ we have
$\nrm{\tB} \leq M^\circ + \frsm{1}{2} (M-M^\circ)$ and therefore $\xi(\tB(t)) = \tB(t)$.
So for $n$ sufficiently large, $\tB$ solves $B = \Upsilon_n(B)$, or, in other words $\tB = \hB$.
We have thus shown that $\nrm{\hB - B^\circ} \cas 0$, as desired.

\nothere{
Define $\tilde G_i=G_i-\mu(L_i;\theta_0)$, $U=(T,\delta,X,L,G)$, and let 
\begin{align*}
\psi_{B,t}(U)&=\int_0^t\tilde G e^{B(s-)X}\{dN(s)-R(s)XdB(s)\},
\end{align*}
and $\Psi(B)(t)=P\psi_{B,t}$. Note that $\Psi(B_X)(t)=0$.
The empirical version is denoted by 
$$
\Psi_n(B)(t)=\mathbb{P}_n\psi_{B,t}=n^{-1}\sum_i\int_0^t\tilde G_i e^{B(s-)X_i}\{dN_i(s)-R_i(s)X_idB(s)\},
$$
and our estimator, $\hat B_X^{ \theta_0}=B_X(\cdot,\theta_0)$, satisfies $\Psi_n(\hat B_X^{\theta_0})(t)=0$.

Let the parameter space $\mathbb{B}$ be the space of all functions $B$ restricted to $[0,\tau]$ so that $B(t)=\int_0^t\beta(s)\, ds$ and $B$ bounded. 
Note that for $B\in \mathbb{B}$ we have that $B$ is of bounded variation so it can be written as $B=B_1-B_2$ where
$B_1$ and $B_2$ are increasing functions.

We claim that the class of functions 
$$
\mathcal{F}=\{\psi_{B,t}:\; B\in  \mathbb{B}, t\in [0,\tau]\}
$$
is Donsker. To see this we note, for $ B\in  \mathbb{B}$, that
$$\psi_{B,t}(U)=\tilde G-\tilde Ge^{B(t-)X}I(t\leq T)-\tilde Ge^{B(T-)X}I( T\leq t)(1-\delta),$$
and that 
 the class of monotone functions $f:\; [0,\tau]\rightarrow [0,1]$ of the real random variable $T$ is Donsker (Kosorok, Thm. 9.24). Also, sums and products of bounded Donsker classes are also Donsker. To show that $
\mathcal{F}$ is Donsker, the only non-trivial thing to show is that the class
$$
\mathcal{G}=\{\tilde\psi_{B,t}:\; B\in  \mathbb{B}, t\in [0,\tau]\},
$$
with $\tilde\psi_{B,t}(U)=e^{B(T-)X}I( T\leq t)$, is Donsker. We assume without loss of generality that $X>0$. Write $\tilde\psi_{B,t}$ as
$$
\tilde\psi_{B,t}(U)=\frac{e^{B_1(T-)X}}{e^{B_1(\tau)X}}e^{-B_2(T-)X}I( T\leq t)e^{B_1(\tau)X}
$$
Use (Kosorok, Thm. 9.24) to conclude that $\mathcal{G}$ is Donsker. Since Donsker classes are also Glivenco-Cantelli, we have that 
$$
\sup_{B\in  \mathbb{B}}|| \Psi_n(B)-\Psi(B)||\cp 0.
$$
We now turn to the identifiability condition (Kosorok, Thm. 2.10). Let 
$$
A_0(t,c)=n^{-1}\sum_{i=1}^n\tilde G_iR_i(t)e^{cX_i}X_i,
$$
then it is clear that we have unif. (in t)  conv. in prob. towards  
$$
a_0(t,c)=E[\tilde G_iR_i(t)e^{cX_i}X_i].
$$
Denote, for $B\in  \mathbb{B}$,
$$
\Upsilon(t,B)=\int_0^t\frac{E(\tilde Ge^{B(s-)X}dN(s))}{a_0(t,B(s-))}
$$
and note that $\Psi(B)(t)=0$ if and only if $\Upsilon(t,B)=B(t)$. We now show that $B_X(t)$ is the unique solution to $\Upsilon(t,B)=B(t)$ following (Zucker, 2005). Denote 
$$
q(s,c)=\frac{E\{\tilde Ge^{cX}\lambda(s|L,G,X)\}}{a_0(t,c)},
$$
where $\lambda(s|L,G,X)ds=E(dN(s)|L,G,X)$.
We shall assume that 
$$
\inf_{t,c}|a_0(t,c)|>0.
$$
 It follows by this condition  that $q$ is bounded over $s$ and $c$, and also that it is Lipschitz continuous with respect to c with a Lipschitz constant independent of $s$. It then follows that there exists a unique solution to $\Upsilon(t,B)=B(t)$ subject to $B(0)=0$.
\nothere{
For known $\theta$ we can write 
$$\hat B_X(t,\theta)=\int_0^tH_{\theta}\{s,\hat B_X(s-,\theta)\}dN(s),
$$
where the $k$th element of the $n$-vector $H_{\theta}\{t,\hat B_X(t-,\theta)\}$ 
is 
$$\{G_k-\mu(L_k;\theta)\}e^{\hat B_X(t-,\theta)X_k}/\sum_{i=1}^n\{G_i-\mu(L_i;\theta)\}R_i(s)e^{\hat B_X(t-,\theta)X_i}X_i.
$$
We first concentrate on showing that $\hat B_X(t,\theta_0)$ is consistent. For any function $\tilde B_X(t)$ define
\begin{align*}
g_n(\tilde B_X)(t)&=\int_0^tH_{\theta_0}\{s,\tilde B_X(s-)\}dN(s)\\
&=B_X(t)+\int_0^tH_{\theta_0}\{s,\tilde B_X(s-)\}\left [dN(s)-X(s)dB_X(s)\right ],
\end{align*}
where $X(t)$ is the $n\times 1$ vector with $i$th entry $R_i(t)X_i$, and note that $\hat B_X(t,\theta_0)$ is fix point, that is,
$$
g_n(\hat B_X(\cdot,\theta_0))(t)=\hat B_X(t,\theta_0)
$$
{\bf Eric and Stijn}: It is tempting to try to mimic the proof of Zucker (JASA, 2005):
Let 
$$
A_0(\theta,t,c)=n^{-1}\sum_{i=1}^n\{G_i-\mu(L_i;\theta)\}R_i(s)e^{cX_i}X_i,
$$
then it is clear that we have unif. (in t)  conv. in prob. towards  
$$
a_0(\theta,t,c)=E[\{G_i-\mu(L_i;\theta)\}R_i(s)e^{cX_i}X_i].
$$
NOTE THAT $E[\{G-\mu(L)\}R(s)e^{cX}X]=E[\{G-\mu(L)\}e^{-\Omega(s,U,L)}E(X|G,L,U)]$. IF NEEDED, IT SEEMS OK TO ME TO ASSUME THAT THIS IS BOUNDED AWAY FROM ZERO, WHICH IS GENERALLY THE CASE IF $E(X|G,L,U)$ DEPENDS ON $G$, WHICH IT DOES  WHEN $G$ IS AN IV.
Let 
$$
\Psi_n(t,\theta,B)=\int_0^tH_{\theta}\{s,B(s-)\}dN(s)
$$
and 
$$
\Psi(t,\theta,B)=\int_0^tE[\{G_i-\mu(L_i;\theta)\}e^{ B(s-)X_i}dN_i(s)/a_0(\theta,s,B(s-)]
$$
Note that $\Psi_n(t,\theta,\hat B_X)=\hat B_X(t,\theta)$ and also that 
$B_X(t,\theta_0)=B_X(t)$ since if we plug $B_X(t)$ into the expression for $\Psi(t,\theta_0,B)$ we get 
\begin{align*}
\Psi(t,\theta_0,B)=&\int_0^tE[\{G_i-\mu(L_i;\theta_0)\}e^{ B_X(s)X_i}dN_i(s)/a_0(\theta_0,s,B_X(s)]\\
=&\int_0^tE[\{G_i-\mu(L_i;\theta)\}e^{B(s)X_i}X_iR_i(s)dB_X(s)/a_0(\theta_0,s,B_X(s)]]\\
=&B_X(t).
\end{align*}
 Then we may try to proceed as in Zucker (2005) showing that 
$$
q_{\theta}(s,c)=E[\{G_i-\mu(L_i;\theta_0)\}e^{B(s)X_i}X_iR_i(s)]\beta_x(s)/a_0(\theta,s,c)
$$
($B_X(t)=\int_0^t\beta_x(s)\, ds$) is (a) bounded in $s$ and $c$ and (b) Lipschitz cont. wrt $c$ with the constant indep. of $s$. This will give us that there exists a unique solution to, $B_X(t,\theta)$, to the equation
$$
\Psi(t,\theta,B)=B.
$$

As in Zucker (2005) one can now define $\tilde B_X^{(n)}(t,\theta)$ as the modified version of 
$\hat B_X(t,\theta)$ defined by linear interpolation between the jumps. If we can now show that
$$
{\cal L}=\{\tilde B_X^{(n)}(t,\theta), n\geq n^{'}\}
$$
is  uniformly bounded and equicontinuous. Given that one can then show
$$
\sup_{t,\theta}|\Psi_n(t,\theta,\tilde B_X^{(n)}(t,\theta))-\Psi(t,\theta,\tilde B_X^{(n)}(t,\theta))|\rightarrow 0\;a.s
$$
using Aalen (1976, Lemma 6.1), and also that
$$
\sup_{t,\theta}|\tilde B_X^{(n)}(t,\theta))-\Psi_n(t,\theta,\tilde B_X^{(n)}(t,\theta))|\rightarrow 0\;a.s
$$
One may then conclude as in Zucker (2005) that any limit point of $\{\tilde B_X^{(n)}(t,\theta)\}$ must satisfy 
$\Psi(t,\theta,B)=B$ but since $B_X(t,\theta)$ is the unique solution to this equation, it is the unique limit point of $\{\tilde B_X^{(n)}(t,\theta)\}$. Now conclude that $\tilde B_X^{(n)}(t,\theta))$, and thus also $\hat B_X^{(n)}(t,\theta))$,  converges a.s uniformly in $t$ and $\theta$ to $B_X(t,\theta)$. 

Hence $\tilde B_X^{(n)}(t,\theta))$ converges a.s. uniformly in $\theta$ and $t$ to $B_X(t,\theta)$, which also hold for $\hat B_X^{(n)}(t,\theta))$
It is also clear 
$$
||g_n(B_X)-B_X ||/R_n\rightarrow 0
$$
in probability as $n\rightarrow \infty$ where $R_n=1$; and also that $Z_n=n^{1/2}\{g_n(B_X)(t)-B_X(t)\}$ converges in distribution towards a Gaussian limit since $Z_n$ is essentially a sum of independent identically zero-mean terms. Hence conditions $(A2)$ and $(C)$ in the Appendix of Martinussen et al. (2002) are fulfilled.
To show the consistency of $\hat B_X(t,\theta_0)$ it only remains to show $(A1)$ in the Appendix of Martinussen et al. (2002)  on 
$$
B_n(B_X,R_n)=\{\tilde B_X:\, || B_X-\tilde B_X||\leq R_n\}.
$$
Write $g_n(\tilde B_X^{(2)})(t)-g_n(\tilde B_X^{(1)})(t)$ as
\begin{align*}
n^{-1}\sum_{i=1}\int_0^t  &\left [ e^{\{\tilde B_X^{(2)}(s-)-B_X(s-)\}X_i}/J_{\theta}\{s,\tilde B_X^{(2)}(s-)\}-
e^{\{\tilde B_X^{(1)}(s-)-B_X(s-)\}X_i}/J_{\theta}\{s,\tilde B_X^{(1)}(s-)\}\right ]\times \\
&\{G_i-\mu(L_i;\theta)\}e^{ B_X(s-)X_i}R_i(s)\left \{dN_i(s)-X_idB_X(s)\right \},
\end{align*}
where 
$$
J_{\theta}\{s,\tilde B_X^{(j)}(t)\}=n^{-1}\sum_{i=1}^n\{G_i-\mu(L_i;\theta)\}R_i(t)e^{\tilde B_X^{(j)}(t-)X_i}X_i,\quad j=1,2.
$$
Since the $X_i$'s are assumed to be bounded and since $\tilde B_X^{(1)}$, $\tilde B_X^{(2)}$ both belongs to 
$
B_n(B_X,R_n)
$
it is seen that $||g_n(\tilde B_X^{(2)})-g_n(\tilde B_X^{(1)})||\rightarrow 0$ in probability as  $n\rightarrow \infty$.
}
}
The consistency of $\hat B_X(t,\hat \theta)$ then follows immediately by a Taylor series expansion since $\hat \theta$ is consistent.
\bigskip
\bigskip
\medskip

\noindent
{\bf Asymptotic normality}
\bigskip
\medskip

\noindent

\noindent
Let $N(t)=\{N_1(t),\ldots N_n(t)\}^T$ and $X=(X_1,\ldots , X_n)$. For known $\theta$ we can write 
$$\hat B_X(t,\theta)=\int_0^tH_{\theta}\{s,\hat B_X(s-,\theta)\}dN(s),
$$
where the $k$th element of the $n$-vector $H_{\theta}\{t,\hat B_X(t-,\theta)\}$ 
is 
$$\{G_k-\mu(L_k;\theta)\}e^{\hat B_X(t-,\theta)X_k}/\sum_{i=1}^n\{G_i-\mu(L_i;\theta)\}R_i(t)e^{\hat B_X(t-,\theta)X_i}X_i.
$$
Let $V(t,\theta)=n^{1/2}\{\hat B_X(t,\theta)- B_X(t)\}$ and let $\dot{H}$ denote the derivative of $H$ with respect to its second argument. It is then easy to see that 
\begin{align*}
V(t,\theta)=&n^{1/2}\int_0^tH(s, B_X(s-))\left [dN(s)-XdB_X(s)\right ]\\
&+\int_0^tV(s-,\theta)\{1+o_p(1)
\}\dot{H}(s,B_X(s-))dN(s) 
\end{align*}
which is a  Volterra-equation, see Andersen et al. (1993), p. 91. The solution to this equation is given by
$$
V(t,\theta)=\int_0^t{\cal F}(s,t)n^{1/2}H(s,B_X(s-))\left [dN(s)-XdB_X(s)\right ]+o_p(1),
$$
where 
$$
{\cal F}(s,t)=\prod_{(s,t]}\left \{1+\dot{H}(\cdot,B_X(\cdot ))dN(\cdot)\right \}
$$
with the latter being a  product integral that converges in probability to some limit. This leads to the iid-representation
$$V(t,\theta)=n^{-1/2}\sum_{i=1}^n\epsilon^B_i(t)$$ with the $\epsilon^B_i(t)$'s being
zero-mean iid terms. Specifically
$$
\epsilon^B_i(t)=\int_0^t{\cal F}(s,t)n^{1/2}\{H(s,B_X(s-))\}_i\left [dN(s)-XdB_X(s)\right ]_i
$$
with $a_i$ being the $i$th element of the vector $a$.
This
together with
\begin{align*}
n^{1/2}\{\hat B_X(t,\hat\theta)-B_X(t)\}&=n^{1/2}\{\hat
B_X(t,\theta)-B_X(t)\}+n^{1/2}\{\hat B_X(t,\hat\theta)-\hat B_X(t,\theta)\}\\
&=n^{1/2}\{\hat B_X(t,\theta)-B_X(t)\}+D_{\theta}(\hat
B_X(t,\theta))_{|\hat\theta}n^{1/2}(\hat\theta-\theta)+o_p(1),
\end{align*}
where $D_{\theta}\{\hat
B_X(t,\theta)\}$ is the first order derivative of $\hat
B_X(t,\theta)$ w.r.t. $\theta$ 
gives an iid-decomposition of $n^{1/2}\{\hat B_X(t,\hat\theta)-B_X(t)\}$:
$$
n^{1/2}\{\hat
B_X(t,\hat\theta)-B_X(t)\}=n^{-1/2}\sum_{i=1}^n\epsilon_i^B(t,\theta)+o_p(1),
$$
where 
\begin{equation}
\label{Eps}
\epsilon_i^B(t,\theta)=\epsilon^B_i(t)+D_{\theta}(\hat
B_X(t,\theta))_{|\theta}\epsilon^{\theta}_i.
\end{equation}
We now  argue that the process $V(t,\theta)$ converges in distribution  as a process using arguments similar to what is done in  Lin et al.  (2000.  p. 726).
By taking the $\log{}$ to equation \eqref{Param} it is seen that  $B_X(t)$ can be written as a difference of two monotone functions .
 Let $\tilde H_i(s)$ be the limit in probability of ${\cal F}(s,t)H_i(s,B_X(s-))$. 
Now, split 	$\tilde H_i(s)$	 into its positive and negative parts, $\tilde H_i^+(s)$ and $\tilde H_i^-(s)$, 
and similarly with $X_i$, $X_i^+$ and $X_i^-$. 
Then $\int_0^t\tilde H_i(s)[ dN_i(s)-X_idB_X(s) ]$ can be written as  a difference of two monotone functions, and then we follow the arguments of Lin et al. (2000) (or use example 2.11.16 of van der Vaart and Wellner, 1996).
Convergence in distribution for the process  $V(t,\hat \theta)$ also holds using the above Taylor expansion.
It thus follows that
$$
n^{1/2}\{\hat B_X(t,\hat\theta)-B_X(t)\}
$$
converges to a zero-mean Gaussian process with a variance that is
consistently estimated by
$$
n^{-1}\sum_{i=1}^n\hat\epsilon_i^B(t,\hat\theta)^2.
$$
The derivative $D_{\theta}(\hat
B_X(t,\theta))_{|\hat\theta}$ can be calculated recursively as $\hat B_X(t,\hat\theta)$ is constant between the observed death times. Denote the jump times by $\tau_1,\ldots ,\tau_m$. Hence
$$
\hat B_X(\tau_j,\theta)=\hat B_X(\tau_{j-1},\theta)+d\hat B_X(\tau_j,\theta)
$$
which then also holds for the derivative. Since $\hat B_X(0,\theta)=0$ and the derivative of the increment in the first jump time,  $d\hat B_X(\tau_1,\theta)$, is easily calculated we then have a recursive way of calculating the derivatives of $\hat B_X(\cdot,\theta)$.


\nothere{
\section*{Appendix: Semi-parametric efficiency}

Let $\mathcal{A}$ be the model for the observed data $(\tilde{T},X,G,L)$ defined by the following restriction:
\[P(\tilde{T}>t|X,G,L)=\exp{\left\{-B_X(t)X\right\}}P(\tilde{T}^0>t|X,G,L)\]
and the additional restriction that $\tilde{T}^0\cip G|L$ and $f(G|L)$ is known. 
Let $\mathcal{B}$ be the model for the observed data $(\tilde{T},X,G,L)$ defined by the following restriction:
\[P(\tilde{T}>t|X,G,L,U)=\exp\left\{-B_X(t)X-\Omega(t,U,L)\right\}\]
with $U$ unobserved, $\Omega(t,U,L)$ unknown, $U\cip G|L$ and $f(G|L)$ known. Model $\mathcal{B}$ satisfies the restrictions of model $\mathcal{A}$. This is easily verified upon setting
\[P(\tilde{T}^0>t|X,G,L)=E\left[\exp\left\{-\Omega(t,U,L)\right\}|X,G,L\right];\]
note indeed that $E\left[\exp\left\{-\Omega(t,U,L)\right\}|G,L\right]$ does not depend on $G$ for all $t$ and $L$, and that $E\left[\exp\left\{-\Omega(t,U,L)\right\}|X,G,L\right]$ is a valid probability since $\exp\left\{-\Omega(t,U,L)\right\}=P(\tilde{T}>t|X=0,G,L,U)$ and thus lies between 0 and 1 for all $t,G,L$ and $U$. In this section, we will first derive the 
orthocomplement of the nuisance tangent space for $B_X(t)$ under model $\mathcal{B}$, which must contain the orthocomplement of the nuisance tangent space
for $B_X(t)$ under model $\mathcal{A}$. We will then verify that all elements in this space are unbiased under model $\mathcal{A}$, from which we can infer that both spaces are identical. 

\subsection*{Orthocomplement of the nuisance tangent space}

ERIC, THE START OF WHAT FOLLOWS BUILDS UPON AN UNPUBLISHED PAPER OF YOURS ON ADDITIVE HAZARD MODELS; WHAT CITATION SHOULD I ADD? The likelihood for a single observation $({T},\Delta,X,G,L,U)$ under model $\mathcal{B}$, evaluated at $(t,\delta,x,g,l,u)$ equals 
\begin{eqnarray*}
&&\lambda_{C|X,G,L,U}(t|x,g,l,u)^{1-\delta}\exp\left\{-\Lambda_{C|X,G,L,U}(t|x,g,l,u)\right\}\lambda_{\tilde{T}|X,G,L,U}(t|x,g,l,u)^{\delta}\\
&&\times \exp\left\{-\Lambda_{\tilde{T}|X,G,L,U}(t|x,g,l,u)\right\} f_{X|G,L,U}(x,g,l,u)f_{G|L}(g,l)f_{U,L}(u,l),
\end{eqnarray*}
where $\lambda_{\tilde{T}|X,G,L,U}(t|x,g,l,u)=d\Omega(t,u,l)+(dB_X(t)/dt)x$ for some $\Omega(t,u,l)$. 
It follows from Bickel et al. (1993) (see also Tsiatis (2006)) that the nuisance tangent space, i.e. the mean square closure of the linear span of score functions for all parametric submodels, is given by 
\[\Lambda_{\rm nuis}=\Lambda_{C|X,G,L,U}\oplus \Lambda_{T|X,G,L,U}\oplus \Lambda_{X,G,L,U}\]
where 
\begin{eqnarray*}
\Lambda_{C|X,G,L,U}&=&\left\{\int x_1(s,X,G,L,U)dM_C(s,X,G,L,U):x_1(s,X,G,L,U) \ \mbox{\rm unrestricted}\right\}\\&&\cap L_2\\
\Lambda_{X,G,L,U}&=&\left\{x_2(X,G,L,U)-E\left\{x_2(X,G,L,U)|G,L,U\right\}+x_2'(U,L):\right.\\&&\left.x_2(X,G,L,U),x_2'(U,L) \ \mbox{\rm unrestricted}\right\}\cap L_2
\end{eqnarray*}
where $M_C(s,X,G,L,U)$ is the martingale corresponding to the counting process $N_C(s)\equiv I(T\leq s,\delta=0)$ w.r.t. the history spanned by ${N}_C(s),X,G,U$ and $L$, and where $L_2$ denotes the Hilbert space of all mean zero functions of the observed data with finite variance.
Under the independent censoring assumption that $C\cip \tilde{T}|X,G,L,U$, we have that $\Lambda_{C|X,G,L,U}^{\perp}$ equals
\begin{eqnarray*}
&&\left\{\int x_3(s,X,G,L,U)dM_T(s,X,G,L,U)+R(s)x_3'(s,X,G,L,U)ds+x_3''(X,G,L,U):\right.\\&&\left.x_3(s,X,G,L,U),x_3'(s,X,G,L,U),x_3''(X,G,L,U) \ \mbox{\rm unrestricted}\right\}\cap L_2,
\end{eqnarray*}
where $M_T(s,X,G,L,U)$ is the martingale corresponding to the counting process $N(s)$ w.r.t. the history spanned by ${N}(s),X,G,U$ and $L$.
The orthocomplement $\left\{\Lambda_{C|X,G,L,U}\oplus \Lambda_{X,G,L,U}\right\}^{\perp}$ consists of all elements of $\Lambda_{C|X,G,L,U}^{\perp}$ that are orthogonal to all nuisance scores in $\Lambda_{X,G,L,U}$. Note that $\int x_3(s,X,G,L,U)dM_T(s,X,G,L,U)$ is orthogonal to these nuisance scores.
We thus require in particular that $x_3'(s,X,G,L,U)$ is such that 
\begin{eqnarray*}
0&=&E\left(\int R(s)x_3'(s,X,G,L,U)ds\left[\left\{x_2(X,G,L,U)-E\left\{x_2(X,G,L,U)|G,L,U\right\}+x_2'(U,L)\right]\right)\\
&=&E\left(\int P(C\geq s|L)\exp\left\{-B_X(s)X-\Omega(s,L,U)\right\}x_3'(s,X,G,L,U)ds\right.\\
&&\left.\times\left[\left\{x_2(X,G,L,U)-E\left\{x_2(X,G,L,U)|G,L,U\right\}+x_2'(U,L)\right]\right),
\end{eqnarray*}

for all $x_2(X,G,L,U)$ and $x_2'(U,L)$, where we use that $P(C\geq s|X,G,U,L)=P(C\geq s|U,L)$ as we assume throughout. It follows that $x_3'(s,X,G,L,U)$ must be of the form
\[\exp\left\{B_X(s)X\right\}\left[x'_4(s,G,L,U)-E\left\{x'_4(s,G,L,U)|L,U\right\}\right].\]
Inferring similar restrictions on $x_3''(s,X,G,L,U)$, we thus obtain that $\left\{\Lambda_{C|X,G,L,U}\oplus \Lambda_{X,G,L,U}\right\}^{\perp}$ equals
\begin{eqnarray*}
&&\left\{\int x_4(s,X,G,L,U)dM_T(s,X,G,L,U)+R(s)\exp\left\{B_X(s)X\right\}
\left[x'_4(s,G,L,U)\right.\right.\\&&\left.\left.
-E\left\{x'_4(s,G,L,U)|L,U\right\}\right]ds+x''_4(G,L,U)-E\left\{x''_4(G,L,U)|L,U\right\}:\right.\\&&\left.
x_4(s,X,G,L,U),x_4'(s,G,L,U),x_4''(G,L,U) \ \mbox{\rm unrestricted}\right\}\cap L_2.
\end{eqnarray*}
To determine the nuisance tangent space $\Lambda_{T|X,G,L,U}$
corresponding to the nuisance parameters indexing $\Omega(t,u,l)$, note that the nuisance scores must be of the form
\[\frac{\delta x_5(t,U,L)}{\lambda_{\tilde{T}|X,G,L,U}(t|X,G,L,U)}-\int_0^t x_5(s,U,L)ds\]
for some $x_5(t,U,L)$. It follows that 
\[\Lambda_{T|X,G,L,U}=\left\{\int \frac{x_5(s,L,U)dM_T(s,X,G,L,U)}{\lambda_{\tilde{T}|X,G,L,U}(s|X,G,L,U)}:x_5(s,L,U) \ \mbox{\rm unrestricted}\right\}\cap L_2.\]
The orthocomplement $\Lambda_{\rm nuis}^{\perp}$ thus consists of all elements of $\left\{\Lambda_{C|X,G,L,U}\oplus \Lambda_{X,G,L,U}\right\}^{\perp}$ that are orthogonal to all nuisance scores in 
$\Lambda_{T|X,G,L,U}$. Note therefore that 
\begin{eqnarray*}
0&=&E\left[\int \frac{x_5(s,L,U)dM_T(s,X,G,L,U)}{\lambda_{\tilde{T}|X,G,L,U}(s|X,G,L,U)}\int x_4(s,X,G,L,U)dM_T(s,X,G,L,U)\right]\\
&=&E\left[\int \frac{x_5(s,L,U)}{\lambda_{\tilde{T}|X,G,L,U}(s|X,G,L,U)}x_4(s,X,G,L,U)\lambda_{\tilde{T}|X,G,L,U}(s|X,G,L,U)R(s)ds\right]\\
&=&E\left[\int x_5(s,L,U)x_4(s,X,G,L,U)R(s)ds\right].
\end{eqnarray*}
Further,
\begin{eqnarray*}
&&E\left[\int\frac{x_5(s,L,U)dM_T(s,X,G,L,U)}{\lambda_{\tilde{T}|X,G,L,U}(s|X,G,L,U)}\right.\\&&\left.\times \int R(t)\exp\left\{B_X(t)X\right\}
\left[x'_4(t,G,L,U)
-E\left\{x'_4(t,G,L,U)|L,U\right\}\right]dt\right]\\
&&=E\left[\int \frac{x_5(s,L,U)\left\{0-\lambda_{\tilde{T}|X,G,L,U}(s|X,G,L,U)R(s)ds\right\}}{\lambda_{\tilde{T}|X,G,L,U}(s|X,G,L,U)}\right.\\&&\left.\times \int_s^{\infty} R(t)\exp\left\{B_X(t)X\right\}
\left[x'_4(t,G,L,U)
-E\left\{x'_4(t,G,L,U)|L,U\right\}\right]dt\right]\\
&&=-E\left[\int\int_s^{\infty} x_5(s,L,U)ds R(t)\exp\left\{B_X(t)X\right\}
\left[x'_4(t,G,L,U)
-E\left\{x'_4(t,G,L,U)|L,U\right\}\right]dt\right]\\&&=0,
\end{eqnarray*}
and likewise all nuisance scores in 
$\Lambda_{T|X,G,L,U}$ are orthogonal to the components of the elements of $\left\{\Lambda_{C|X,G,L,U}\oplus \Lambda_{X,G,L,U}\right\}^{\perp}$ that 
involve $x''_4(G,L,U)$. It can thus be inferred that $x_4(s,X,G,L,U)$ must be of the form
\[x_6(s,X,G,L,U)-E\left\{x_6(s,X,G,L,U)|T\geq s,L,U\right\},\]
from which
\begin{eqnarray*}
\Lambda_{\rm nuis}^{\perp}&=&\left\{\int \left[x_6(s,X,G,L,U)-E\left\{x_6(s,X,G,L,U)|T\geq s,L,U\right\}\right]dM_T(s,X,G,L,U)\right.\\
&&\left.+R(s)\exp\left\{B_X(s)X\right\}\left[x'_6(s,G,L,U)-E\left\{x'_6(s,G,L,U)|L,U\right\}\right]ds+ x_6'(G,L,U)\right.\\
&&\left.-E\left\{x_6'(G,L,U)|L,U\right\}:x_6(s,X,G,L,U),x_6'(s,G,L,U),x_6''(G,L,U) \ \mbox{\rm unrestricted}\right\}.
\end{eqnarray*}

The orthocomplement of the nuisance tangent space in the observed data model consists of all functions of the observed data whose conditional expectation, given the full data $(C,T,X,G,L,U)$ belongs to 
$\Lambda_{\rm nuis}^{\perp}$. From the identity
\[E\left\{x_6(s,X,G,L,U)|T\geq s,L,U\right\}=\frac{E\left[x_6(s,X,G,L,U)\exp\left\{-B_X(s)X\right\}|L,U\right]}{E\left[\exp\left\{-B_X(s)X\right\}|L,U\right]},\]
it is seen that the only such functions are obtained by choosing $x_6(s,X,G,L,U)=\exp\left\{B_X(s)X\right\}\left[x_7(s,G,L)-E\left\{x_7(s,G,L)|L\right\}\right]+\tilde{x}_7(s,L),x_6'(s,G,L,U)=x_7'(s,G,L)+d\Omega(s,U,L)$ and $x_6''(G,L,U)=x_7''(G,L)$. The 
 element of $\Lambda_{\rm nuis}^{\perp}$ corresponding to this choice is
\begin{eqnarray*}
&&\int R(s)\exp\left\{B_X(s)X\right\}\left[x_7(s,G,L)-E\left\{x_7(s,G,L)|L\right\}\right]\left\{dN(s)-dB_X(s)X\right\}\\
&&+R(s)\exp\left\{B_X(s)X\right\}\left[x'_7(s,G,L)-E\left\{x'_7(s,G,L)|L\right\}\right]ds+ x_7''(G,L)-E\left\{x_7''(G,L)|L\right\}.
\end{eqnarray*}
With the following notation, $R_0(s)\equiv R(s)\exp\left\{B_X(s)X\right\}$ and $dN_0(s)\equiv dN(s)-dB_X(s)X$, we thus obtain the following orthocomplement:
\begin{eqnarray*}
&&\left\{\int R_0(s)\left(\left[x_7(s,G,L)-E\left\{x_7(s,G,L)|L\right\}\right]dN_0(s)+\left[x'_7(s,G,L)-E\left\{x'_7(s,G,L)|L\right\}\right]ds\right)\right.\\
&&\left. + x_7''(G,L)-E\left\{x_7''(G,L)|L\right\}:
x_7(s,G,L),x_7'(s,G,L),x''_7(G,L) \ \mbox{\rm unrestricted}\right\}.
\end{eqnarray*}
Note that these estimating functions are all unbiased under the less restrictive model $\mathcal{A}$ with the additional 
 independent censoring assumption that $C\cip (\tilde{T},X,G)|L,U$ for some variable $U$ such that $U\cip G|L$.
We thus conclude that the above space also defines the orthocomplement of the nuisance tangent space under this less restrictive model.

\subsection*{Efficient score}

We will derive the efficient score for $B_X(t)$ under model $\mathcal{B}$ with the additional 
 independent censoring assumption that $C\cip (\tilde{T},X,G)|L,U$. It follows from the reasoning in the previous paragraph that the result is also the efficient score under  
 $\mathcal{A}$ (with this additional assumption).

Consider a one-dimensional parameterization $B_X(t;\beta)$ of $B_X(t)$ so that $\partial \left\{dB_X(t;\beta)/dt\right\}/\partial\beta=b(t)$ for some function $b(t)$. 
Then, by a similar reasoning as before, the score $S_{\beta}$ for $\beta$ in the corresponding submodel is given by
\[E\left[\int \frac{Xb(t)}{\lambda_{\tilde{T}|X,G,L,U}(t|X,G,L,U)}dM_T(s,X,G,L,U)|T,X,G,L\right].\]
Being an element of the orthocomplement $\Lambda_{\rm nuis}^{\perp}$, the efficient score for $\beta$ in model $\mathcal{B}$ is of the form
\begin{eqnarray*}
&&\int R_0(s)\left[x^{\rm eff}_6(s,G,L)-E\left\{x^{\rm eff}_6(s,G,L)|L\right\}\right]dN_0(s)\\&&+R_0(s)\left[x^{\rm eff '}_6(s,G,L)-E\left\{x^{\rm eff '}_6(s,G,L)|L\right\}\right]ds+x^{\rm eff ''}_6(G,L)-E\left\{x^{\rm eff ''}_6(G,L)|L\right\}\end{eqnarray*}
for certain functions $x^{\rm eff}_6(s,G,L),x^{\rm eff '}_6(G,L)$ and $x^{\rm eff ''}_6(G,L)$. Since it equals the projection of $S_{\beta}$ onto the orthocomplement of the nuisance tangent space, we must have that for all $x''_6(G,L)$:
\begin{eqnarray*}
0&=&E\left(\left[\int \frac{Xb(t)dM_T(t,X,G,L,U)}{\lambda_{\tilde{T}|X,G,L,U}(t|X,G,L,U)}-R_0(t)dN_0(t)\left[x^{\rm eff}_6(t,G,L)-E\left\{x^{\rm eff}_6(t,G,L)|L\right\}\right]\right.
\right.\\
&&\left.\left.
-R_0(t)\left[x^{\rm eff '}_6(t,G,L)-E\left\{x^{\rm eff '}_6(t,G,L)|L\right\}\right]dt-x^{\rm eff '}_6(G,L)+E\left\{x^{\rm eff '}_6(G,L)|L\right\}\right]
\right.\\
&&\left.\times\left\{x''_6(G,L)-E\left\{x''_6(G,L)|L\right\}\right\}\right)
\end{eqnarray*}
from which $x^{\rm eff '}_6(G,L)$ equals
\begin{eqnarray*}
&&-\int \left[x^{\rm eff}_6(t,G,L)-E\left\{x^{\rm eff}_6(t,G,L)|L\right\}\right]E\left\{R_0(t)dN_0(t)|G,L\right\}\\
&&+\left[x^{\rm eff '}_6(t,G,L)-E\left\{x^{\rm eff '}_6(t,G,L)|L\right\}\right]E\left\{R_0(t)|G,L\right\}dt\\
&&=-\int \left[x^{\rm eff}_6(t,G,L)-E\left\{x^{\rm eff}_6(t,G,L)|L\right\}\right]E\left\{R_0(t)dN_0(t)|L\right\}\\
&&+\left[x^{\rm eff '}_6(t,G,L)-E\left\{x^{\rm eff '}_6(t,G,L)|L\right\}\right]E\left\{R_0(t)|L\right\}dt.
\end{eqnarray*}
It thus follows that the efficient score must be of the form
\begin{eqnarray*}
&&\int \left[x^{\rm eff}_6(s,G,L)-E\left\{x^{\rm eff}_6(s,G,L)|L\right\}\right]\left[R_0(s)dN_0(s)-E\left\{R_0(s)dN_0(s)|L\right\}\right]
\\&&+\left[x^{\rm eff '}_6(s,G,L)-E\left\{x^{\rm eff '}_6(s,G,L)|L\right\}\right]\left[R_0(s)-E\left\{R_0(s)|L\right\}\right]ds.\end{eqnarray*}
We further need that for all $x'_6(s,G,L)$:
\begin{eqnarray*}
0&=&E\left\{\left(\int \frac{Xb(t)dM_T(t,X,G,L,U)}{\lambda_{\tilde{T}|X,G,L,U}(t|X,G,L,U)}
\right.\right.\\&&\left.\left.
-\left[x^{\rm eff}_6(t,G,L)-E\left\{x^{\rm eff}_6(t,G,L)|L\right\}\right]\left[R_0(t)dN_0(t)-E\left\{R_0(t)dN_0(t)|L\right\}\right]
\right.\right.\\&&\left.\left.
-\left[x^{\rm eff '}_6(t,G,L)-E\left\{x^{\rm eff '}_6(t,G,L)|L\right\}\right]\left[R_0(t)-E\left\{R_0(t)|L\right\}\right]dt\right)
\right.\\
&&\left. \times \int \left[x'_6(s,G,L)-E\left\{x'_6(s,G,L)|L\right\}\right]R_0(s)ds\right\}.
\end{eqnarray*}
Using a similar reasoning as before, it can be show that the first term of this expression
\[E\left\{\int \frac{Xb(t)dM_T(t,X,G,L,U)}{\lambda_{\tilde{T}|X,G,L,U}(t|X,G,L,U)}\int \left[x'_6(s,G,L)-E\left\{x'_6(s,G,L)|L\right\}\right]R_0(s)ds\right\}=0\]
for all $x'_6(s,G,L)$. The second term of this expression is
\begin{eqnarray*}
&&E\left\{\int\left[x^{\rm eff}_6(t,G,L)-E\left\{x^{\rm eff}_6(t,G,L)|L\right\}\right]\left[R_0(t)dN_0(t)-E\left\{R_0(t)dN_0(t)|L\right\}\right]
\right.\\&&\left.
\times \int \left[x'_6(s,G,L)-E\left\{x'_6(s,G,L)|L\right\}\right]R_0(s)ds\right\}\\
&&=E\left\{\int\left[x^{\rm eff}_6(t,G,L)-E\left\{x^{\rm eff}_6(t,G,L)|L\right\}\right]\left[x'_6(s,G,L)-E\left\{x'_6(s,G,L)|L\right\}\right]
\right.\\&&\left.
\times \left[R_0(\max(s,t))dN_0(t)-E\left\{R_0(t)dN_0(t)|L\right\}E\left\{R_0(s)|L\right\}\right]ds\right\}.
\end{eqnarray*}
In general, this integral equation has no simple solution, but simplifies under the null hypothesis that $B_{X}(t)=0$ for all $t$ to 
\[\int E\left\{\left[x^{\rm eff}_6(t,G,L)-E\left\{x^{\rm eff}_6(t,G,L)|L\right\}\right]A_1(s,t,L)x'_6(s,G,L)ds\right\}\]
where 
\[A_1(s,t,L)\equiv E\left\{R_0(\max(s,t))dN_0(t)|L\}-E\left\{R_0(t)dN_0(t)|L\right\}E\left\{R_0(s)|L\right\}\right].\]
This can be understood upon noting that 
\begin{eqnarray*}
E\left\{R_0(\max(s,t))dN_0(t)|G,L\}&=&I(s\leq t)E\left\{R_0(t)dN_0(t)|L\right\}\\&&+I(s>t)E\left[R_0(s)\left\{0-dB_X(s)X\right\}|G,L\right]\\
&=&I(s\leq t)E\left\{R_0(t)dN_0(t)|L\right\}
\end{eqnarray*}
when $B_{X}(t)=0$ for all $t$.
The third term of this expression is
\begin{eqnarray*}
&&E\left\{\int\left[x^{\rm eff '}_6(t,G,L)-E\left\{x^{\rm eff '}_6(t,G,L)|L\right\}\right]\left[R_0(t)-E\left\{R_0(t)|L\right\}\right]dt
\right.\\&&\left.
\times \int \left[x'_6(s,G,L)-E\left\{x'_6(s,G,L)|L\right\}\right]R_0(s)ds\right\}\\
&&=E\left\{\int\left[x^{\rm eff '}_6(t,G,L)-E\left\{x^{\rm eff '}_6(t,G,L)|L\right\}\right]\left[x'_6(s,G,L)-E\left\{x'_6(s,G,L)|L\right\}\right]
\right.\\&&\left.
\times \left[R_0(\max(s,t))-E\left\{R_0(t)|L\right\}E\left\{R_0(s)|L\right\}\right]dsdt\right\}\\
&&=\int E\left\{\left[x^{\rm eff '}_6(t,G,L)-E\left\{x^{\rm eff '}_6(t,G,L)|L\right\}\right]A_2(s,t,L)x'_6(s,G,L)dsdt\right\},
\end{eqnarray*}
where 
\[A_2(s,t,L)\equiv E\left\{R_0(\max(s,t))|L\}-E\left\{R_0(t)|L\right\}E\left\{R_0(s)|L\right\}\right].\]
Combining these results, we thus find that $x^{\rm eff '}_6(t,G,L)$ must solve the following integral equation:
\begin{eqnarray*}
0&=&\int \left[x^{\rm eff}_6(t,G,L)-E\left\{x^{\rm eff}_6(t,G,L)|L\right\}\right]A_1(s,t,L)dt\\&&+\left[x^{\rm eff '}_6(t,G,L)-E\left\{x^{\rm eff '}_6(t,G,L)|L\right\}\right]A_2(s,t,L)dt
\end{eqnarray*}
for all $s$. It follows that 
\[x^{\rm eff '}_6(t,G,L)-E\left\{x^{\rm eff '}_6(t,G,L)|L\right\}=\int \left[x^{\rm eff}_6(u,G,L)-E\left\{x^{\rm eff}_6(u,G,L)|L\right\}\right]B(u,t,L)du\]
for some function $B(u,t,L)$, which can be solved from the set of integral equations:
\[A_1(s,u,L)+\int B(u,t,L)A_2(s,t,L)dt=0\]
for all $s,u$. We thus conclude that when $B_X(t)=0$ for all $t$, the efficient score must be of the form
\begin{eqnarray*}
&&\int \left[x^{\rm eff}_6(s,G,L)-E\left\{x^{\rm eff}_6(s,G,L)|L\right\}\right]\left[R_0(s)dN_0(s)-E\left\{R_0(s)dN_0(s)|L\right\}\right]
\\&&+\int \left[x^{\rm eff}_6(u,G,L)-E\left\{x^{\rm eff}_6(u,G,L)|L\right\}\right]B(u,s,L)du\left[R_0(s)-E\left\{R_0(s)|L\right\}\right]ds\\
&&=\int \left[x^{\rm eff}_6(s,G,L)-E\left\{x^{\rm eff}_6(s,G,L)|L\right\}\right]\left(R_0(s)dN_0(s)-E\left\{R_0(s)dN_0(s)|L\right\}\right.
\\&&\left.+\int B(s,u,L)\left[R_0(u)-E\left\{R_0(u)|L\right\}\right]du\right)ds.\end{eqnarray*}

Finally, we need that for all $x_6(s,G,L)$:
\begin{eqnarray*}
0&=&E\left\{\left(\int \frac{Xb(t)dM_T(t,X,G,L,U)}{\lambda_{\tilde{T}|X,G,L,U}(t|X,G,L,U)}
\right.\right.\\&&\left.\left.
-\left[x^{\rm eff}_6(t,G,L)-E\left\{x^{\rm eff}_6(t,G,L)|L\right\}\right]\left(R_0(t)dN_0(t)-E\left\{R_0(t)dN_0(t)|L\right\}\right.
\right.\right.\\&&\left.\left.\left.
+\int B(s,u,L)\left[R_0(u)-E\left\{R_0(u)|L\right\}\right]du\right)\right.
\right.\\
&&\left. \times \int \left[x_6(s,G,L)-E\left\{x_6(s,G,L)|L\right\}\right]R_0(s)dN_0(s)ds\right\}.
\end{eqnarray*}
Unfortunately, also this integral equation admits no simple solution, which I demonstrate next.
The first term reduces to 
\begin{eqnarray*}
&&E\left\{\int \frac{Xb(t)dM_T(t,X,G,L,U)}{\lambda_{\tilde{T}|X,G,L,U}(t|X,G,L,U)}\int \left[x_6(s,G,L)-E\left\{x_6(s,G,L)|L\right\}\right]\right.\\
&&\left. \times \exp\left\{B_X(s)X\right\}\left\{dM_T(s,X,G,L,U)+R(s)d\Omega(s,U,L)\right\}\right\}\\
&&=E\left\{\int Xb(t)\left[x_6(t,G,L)-E\left\{x_6(t,G,L)|L\right\}\right] \exp\left\{B_X(t)X\right\}dt \right.\\
&&\left.+\int\int I(s>t)\left\{-Xb(t)dt\right\}\left[x_6(s,G,L)-E\left\{x_6(s,G,L)|L\right\}\right]R_0(s)dN_0(s)ds\right\}
\end{eqnarray*}
Under the null hypothesis, the second term reduces to 
\begin{eqnarray*}
&&E\left\{\int \left[x^{\rm eff}_6(t,G,L)-E\left\{x^{\rm eff}_6(t,G,L)|L\right\}\right]\left[x_6(t,G,L)-E\left\{x_6(t,G,L)|L\right\}\right]R(t)dN^2(t)
\right.\\
&&\left.-\int\int  \left[x^{\rm eff}_6(t,G,L)-E\left\{x^{\rm eff}_6(t,G,L)|L\right\}\right]\left[x_6(s,G,L)-E\left\{x_6(s,G,L)|L\right\}\right]E\left\{R(t)dN(t)|L\right\}E\left\{R(s)dN(s)|L\right\}\right\}\\
&&=E\left\{\int \left[x^{\rm eff}_6(t,G,L)-E\left\{x^{\rm eff}_6(t,G,L)|L\right\}\right]x_6(t,G,L)E\left\{R(t)dN(t)|L\right\}
\right.\\
&&\left.-\int\int  \left[x^{\rm eff}_6(t,G,L)-E\left\{x^{\rm eff}_6(t,G,L)|L\right\}\right]x_6(s,G,L)E\left\{R(t)dN(t)|L\right\}E\left\{R(s)dN(s)|L\right\}\right\}\\
&&=\int E\left\{\left[x^{\rm eff}_6(t,G,L)-E\left\{x^{\rm eff}_6(t,G,L)|L\right\}\right]E\left\{R(t)dN(t)|L\right\}\right.\\&&\left.\times\left[x_6(t,G,L)-\int x_6(s,G,L)E\left\{R(s)dN(s)|L\right\}ds\right]\right\}.
\end{eqnarray*}
The third term reduces to 
\begin{eqnarray*}
&&\int\int E\left\{\left[x^{\rm eff}_6(t,G,L)-E\left\{x^{\rm eff}_6(t,G,L)|L\right\}\right]\int B(s,u,L)A_1(s,u,L)du x_6(s,G,L)dsdt\right\}.
\end{eqnarray*}
}

\section*{References}

\noindent Aalen, O. (1976). Nonparametric Inference in Connection with Multiple Decrement
Models. \textit{Scandinavian Journal of Statistics}, \textbf{3}, 15-27.
\medskip

\noindent Aalen, O. O. (1980). A model for non-parametric regression analysis of counting processes. \textit{Lecture Notes in Statistics}, {\bf 2}, 1-25.
 \medskip
 
 \noindent Abadie, A. (2003). Semiparametric instrumental variable estimation of treatment response models \textit{journal of Econometrics}, {\bf 113}, 231-263.
 \medskip

\noindent Andersen, P. K., and Gill, R. D. (1982). Cox's Regression Model for Counting
Processes: A Large Sample Study.  \textit{Annals of Statistics}, \textbf{10}, 1100-1120.
\medskip

\noindent Andersen, P. K., Borgan, O., Gill, R. D. and
 Keiding, N. (1993). \textit{Statistical Models Based on Counting Processes}. Berlin: Springer-Verlag. 
\medskip


\noindent	Angrist, J. and Imbens, G. (1991). 
Sources of identifying information in evaluation models. 
Technical Working Paper 117, National Bureau of Economic Research, Cambridge, MA.
 \medskip

\noindent	Angrist, J. and Krueger, A. (2001). Instrumental variables and the search for identification: From supply and demand to natural experiments. \textit{Journal of Economic Perspectives} \textbf{15}, 69-85.
 \medskip
 
\noindent Bochud, M.  and  Rousson, V. (2010).
 Usefulness of mendelian randomization in observational epidemiology
  \textit{Int. J. Environ. Res. Public Health}, \textbf{7}, 4726-4747.
\medskip

 \noindent Boef, A.G. C.,  le Cessie, S. and Dekkers, O. M.  (2015).
Mendelian 
randomization studies 
in the elderly.
 \textit{Epidemiology}, \textbf{26}, e15-e16.
 \medskip
 
  \noindent Burgess, S. and CRP CHD Genetics Collaboration (2013). Identifying the odds ratio estimated by a two-stage instrumental variable analysis with logistic regression model. \textit{Statistics in Medicine}, \textbf{32}, 711-728.
\medskip

 \noindent Cai, B., Small, D. S. and Ten Have, T. R. (2011). Two-stage instrumental variable methods for estimating the causal odds ratio: analysis of bias.
\textit{Statistics in Medicine}, \textbf{30}, 1809-1824.
\medskip

  \noindent Clarke, P. S. and Windmeijer, F. (2010). Identification of causal effects on binary outcomes using structural mean models.
\textit{Biostatistics}, \textbf{11}, 756-770.
\medskip

  \noindent Clarke, P. S. and Windmeijer, F. (2012). Instrumental variable estimators for binary outcomes.
\textit{Journal of the American Statistical Association}, \textbf{107}, 1638-1652.
\medskip
 
 \noindent  Coddington, E. A. (1989). \textit{An Introduction to Ordinary Differential Equations}. Mineola: Dover
\medskip

\noindent Cuzick, J., Sasieni, P., Myles, J., et al. (2007). Estimating the effect of treatment in a proportional hazards model in the presence of non-compliance and contamination. \textit{Journal of the Royal Statistical Society - Series B} {\bf 69}, 565-588.
 \medskip
 
\noindent Davey-Smith, G. and Ebrahim, S. (2003). Mendelian randomization': can genetic epidemiology contribute to understanding environmental determinants of disease?
\textit{International Journal of Epidemiology} {\bf 32}, 1-22.
 \medskip

\noindent Didelez, V. and Sheehan, N. (2007). Mendelian randomization as an instrumental variable approach to causal inference. \textit{Statistical Methods in Medical Research} {\bf 16}, 309-330.
 \medskip
 
\noindent  
Hartman, P. (1973). \textit{Ordinary Differential Equations}, 2nd ed. (reprinted, 1982),
Boston: Birkhauser.
\medskip

  \noindent Harbord, R. M., Didelez, V., Palmer, T. M., Meng, S., Sterne, J. A. C and Sheehan, N. A. (2012). Severity of bias of a simple estimator of the causal odds ratio in Mendelian randomization studies.
\textit{Statistics in Medicine}, \textbf{32}, 1246-1258.
\medskip

\noindent Hern\'an, M. A. and Robins J. M. (2006). Instruments for causal inference: an epidemiologist's dream?
\textit{Epidemiology} {\bf 17}, 360-372.
\medskip

\noindent	Imbens, G. W. and Angrist, J.  (1994). Identification and estimation of local average treatment effects.
 \textit{Econometrica} \textbf{62}, 467-476.
 \medskip
 
 \noindent  Joffe, M.M. (2001). Administrative and artificial censoring in censored regression models.
 \textit{Statistics in Medicine} {\bf 20}, 2287-2304.
 \medskip
 
\noindent  Joffe, M.M., Yang, W.P. and Feldman, H. (2012). G-Estimation and Artificial Censoring: Problems, Challenges, and Applications. \textit{Biometrics} {\bf 68}, 275-286.
 \medskip
 
\noindent  
 Katan M. B. (1986) Apolipoprotein E isoforms, serum
cholesterol, and cancer. {\it Lancet} 
507-8.
 \medskip
 
\noindent Kosorok, M. R. (2008). \textit{Introduction to Empirical Processes and Semiparametric Inference}. Berlin: Springer-Verlag. 
\medskip

\noindent
Li, J.,  Fine, J.  and  Brookhart, A. (2014).  Instrumental variable additive hazards models.
\textit{Biometrics}, \textbf{71}, 122-130.
 \medskip

\noindent
Lin, D. Y., Wei, L. J., Yang, I. and Ying, Z. (2000). Semiparametric regression for the mean and rate functions of recurrent events. \textit{Journal of the Royal Statistical Society - Series B}, \textbf{62}, 711-730.
\medskip

\noindent Loeys, T., Goetghebeur, E and Vandebosch, A. (2005). Causal proportional hazards models and time-constant exposure in randomized clinical trials. \textit{Lifetime Data Analysis} {\bf 11}, 435-449.
 \medskip

\noindent
Martinussen, T. (2010). Dynamic path analysis for event time data: large sample properties and inference. \textit{ Lifetime Data Analysis}, \textbf{16}, 85-101.
\medskip

\noindent Martinussen, T., Vansteelandt, S., Gerster, M. and Hjelmborg, J. V. B. (2011)
Estimation of direct effects for survival data by using the Aalen additive hazards model.
\textit{Journal of the Royal Statistical Society - Series B}, \textbf{73}, 773-788.
 \medskip

\noindent	Mildner M, Jin J, Eckhart L, Kezic S, Gruber F, Barresi C, et al. (2010).
 Knockdown of filaggrin impairs diffusion barrier function and increases UV sensitivity in a human skin model. 
 \textit{J Invest Dermatol}  \textbf{130}, 2286-2294.
 \medskip

\noindent Nie, H., Cheng, J., and Small, D.S. (2011). Inference for the Effect of Treatment on Survival Probability in Randomized Trials with Noncompliance and Administrative Censoring. \textit{Biometrics} {\bf  67}, 1397-1405.
  \medskip

 \noindent		Olsen MH, Hansen TW, Christensen MK, Gustafsson F, Rasmussen S, Wachtell K, et al. (2007).
 N-terminal pro-brain natriuretic peptide, but not high sensitivity C-reactive protein, improves cardiovascular risk prediction in the general population. \textit{Eur Heart J} 28(11):1374-81.
 \medskip

\noindent	Palmer CN, Irvine AD, Terron-Kwiatkowski A, Zhao Y, Liao H, Lee SP, et al. (2006).
 Common loss-of-function variants of the epidermal barrier protein filaggrin are a major predisposing factor for atopic dermatitis.  \textit{Nat Genet}
\textbf{38}, 441-446.
 \medskip

\noindent
Pearl, J. (2000). \textit{Causality: Models, Reasoning, and Inference}.
Cambridge University Press, Cambridge.
\medskip

\noindent
Picciotto, S., Hern\'an, M. A., Page, J., Young, J. G. and Robins, J. M. (2012). Structural nested cumulative failure time models to estimate the effects of hypothetical interventions. \textit{Journal of the American Statistical Association} {\bf 107}, 886-900.
\medskip



\noindent
Raisin, J. A., Schneeweiss, S. , Glynn, R. J., Mittleman, M. A. and Brookhart, M. A. (2008). 
Instrumental Variable Analysis for Estimation of Treatment Effects With Dichotomous Outcomes.
 \textit{American Journal of Epidemiology} {\bf 169}, 273-284..
\medskip

\noindent
Robins, J.M. and Tsiatis, A. (1991). Correcting for non-compliance in randomized trials using rank-preserving structural failure time models. \textit{Communications in Statistics} {\bf 20}, 2609-2631.
\medskip

\noindent
Robins, J.M. and Rotnitzky, A. (2004). Estimation of treatment effects in randomised trials with non-compliance and a dichotomous outcome using structural mean models. \textit{Biometrika} {\bf 91}, 763-783.
\medskip

\noindent
Shapiro, S.  (1977). Evidence of screening for breast cancer from a randomised trial.
,\textit{Cancer} {\bf 39}, 2772-2782.
\medskip

\nothere{
\noindent
Skaaby, T.,  Husemoen, L. L. N., Martinussen, T., 
Thyssen, J. P., Melgaard, M., et al. (2013).
Vitamin D status , Filaggrin genotype and cardiovascular risk factors: a 
Mendelian randomisation approach.
\textit{PLoS ONE}  8(2): e57647. \\
doi:10.1371/journal.pone.0057647
\medskip
}

\noindent
Tchetgen Tchetgen, E. J., Walter, S., Vansteelandt, S., Martinussen, T., Glymour, M. (2015). Instrumental variable estimation in a survival context.
\textit{ Epidemiology} {\bf26}, 402-410.
\medskip

\noindent
Tsiatis, A. A. (2006).
\textit{Semiparametric Theory and Missing Data}.  Springer Verlag.
\medskip

 \noindent	van den Oord RA, Sheikh A. (2009). Filaggrin gene defects and risk of developing allergic sensitisation and allergic disorders: systematic review and meta-analysis. \textit{BMJ} 339:b2433.
 \medskip

\noindent
Vansteelandt, S. and Goetghebeur, E. (2003). Causal inference with generalized structural mean models. \textit{Journal of the Royal Statistical Society, Series B} \textbf{65}, 817- 835.
\medskip

\noindent
Vansteelandt, S., Bowden, J., Babanezhad, M. and Goetghebeur, E. (2011). On instrumental variable estimation of the causal odds ratio.
\textit{Statistical Science}, \textbf{26}, 403-422.
\medskip

\nothere{
\noindent
Vansteelandt, S. (2009). Estimating direct effects in cohort and case-control studies. \textit{Epidemiology}, \textbf{20}, 851-860.\medskip

\noindent
Vansteelandt, S., Bekaert, M. and Claeskens, G. (2010). On model selection and model misspecification in causal inference. \textit{Statistical Methods in Medical Research}, \newline
doi: 10.1177/0962280210387717.
\medskip

\noindent
Zucker, DM. (2005). A pseudo partial likelihood method for semi-parametric survival regression with covariate errors. \textit{Journal of the American Statistical Association} \textbf{100}, 1264-1277.

}

\end{document}